\documentclass[aps,prl,twocolumn,nopacs,superscriptaddress,footinbib,floatfix]{revtex4}

\usepackage{graphicx}
\usepackage{amsmath}
\usepackage{amssymb}
\usepackage{booktabs}
\usepackage{xcolor}
\usepackage{upgreek}

\usepackage[normalem]{ulem}

\usepackage{bbold}

\usepackage[utf8]{inputenc}

\def\vec #1{{\bf #1}}

\newcommand{\be}{\begin{eqnarray}}
\newcommand{\ee}{\end{eqnarray}}

\renewcommand\vec{\boldsymbol}
\newcommand{\dd}{\mathrm{d}}
\newcommand{\ii}{\mathrm{i}}
\DeclareMathOperator{\Tr}{Tr}

\renewcommand\vec{\boldsymbol}
\newcommand{\quotes}[1]{`#1'}

\begin{document}

\title{Nematic bits and universal logic gates}

\author{\v{Z}iga Kos}
\affiliation{Department of Mathematics, Massachusetts Institute of Technology, 77 Massachusetts Avenue, Cambridge,~MA~02139, USA}
\affiliation{Faculty of Mathematics and Physics, University of Ljubljana, Jadranska 19, 1000 Ljubljana, Slovenia.}
\author{J\"orn Dunkel} 
\affiliation{Department of Mathematics, Massachusetts Institute of Technology, 77 Massachusetts Avenue, Cambridge,~MA~02139, USA}
\date{\today}


\begin{abstract}
Liquid crystals (LCs) can host robust topological defect structures that essentially determine their optical and elastic properties. Although recent experimental progress enables precise control over localization and dynamics of nematic LC defects, their practical potential for information storage and processing has yet to be explored. Here, we introduce  the concept of nematic bits (nbits) by exploiting a quaternionic mapping from LC defects to the Poincar\'e-Bloch sphere.  Through theory and simulations, we demonstrate how single-nbit  operations can be implemented using electric fields, in close analogy with Pauli, Hadamard and other common quantum gates. Ensembles of two-nbit states can exhibit strong statistical correlations arising from nematoelastic interactions, which can be used as a computational resource. Utilizing nematoelastic interactions, we show how suitably arranged 4-nbit configurations can realize universal classical NOR and NAND gates. Finally, we demonstrate the implementation of generalized  logical functions that take values on the Poincar\'e-Bloch sphere. These results open a new route towards the implementation of classical and non-classical computation strategies in topological soft matter systems.
\end{abstract}

\maketitle

Bits are the fundamental units of binary digital computation and information storage. Similar to an idealized universal Turing machine~\cite{1937Turing},  classical digital computers represent bits as two discrete voltage states, commonly labeled 0 and 1. Accordingly, electronic digital circuits process information by manipulating deterministic bit sequences $\cdot\cdot 01011\cdot\cdot$ with the help of logic  gates.
Notwithstanding the historical success of classical bit-based computation, it has long been suggested that some practically relevant problems~\cite{1996Grover,1997DeutschJozsa,1997Shor_SIAM} could be solved by performing parallel computations in larger or non-discrete state spaces~\cite{bournez_2008,PaunG_TheoreticalComputerScience287_2002,BLUML_IntJBifurcationChaos6_2011,AaronsonS_SIGACTNews36_2005,AdlemanL_Science266_1994,WolframS_CommunicationsinMathematicalPhysics96_1984,PhysRevA.58.R1633}. The perhaps best known examples are quantum computers~\cite{1995QuantumGates_PRA,Ladd:2010aa}, which operate on continuous many-qubit Hilbert spaces~\cite{2001Preskill_PRA} and promise substantially faster search~\cite{1996Grover} and factoring~\cite{1997Shor_SIAM} algorithms. Several other information processing strategies in classical systems are currently also being explored, including DNA-based computation~\cite{2011Winfree_DNA}, analog-computing in cells~\cite{Daniel:2013aa}, chemical computers~\cite{2005Adamatzky_Book}, or holonomic computation~\cite{1999Zanardi_PLA} in non-Abelian mechanical~\cite{Fruchart:2020aa} systems. Independent of whether such non-standard approaches will eventually result in scalable computing technologies, their exploration has generally led to a better experimental and theoretical understanding of the underlying physical, chemical and biological systems.

\begin{figure}[b!]
\centering
\includegraphics[width=\columnwidth]{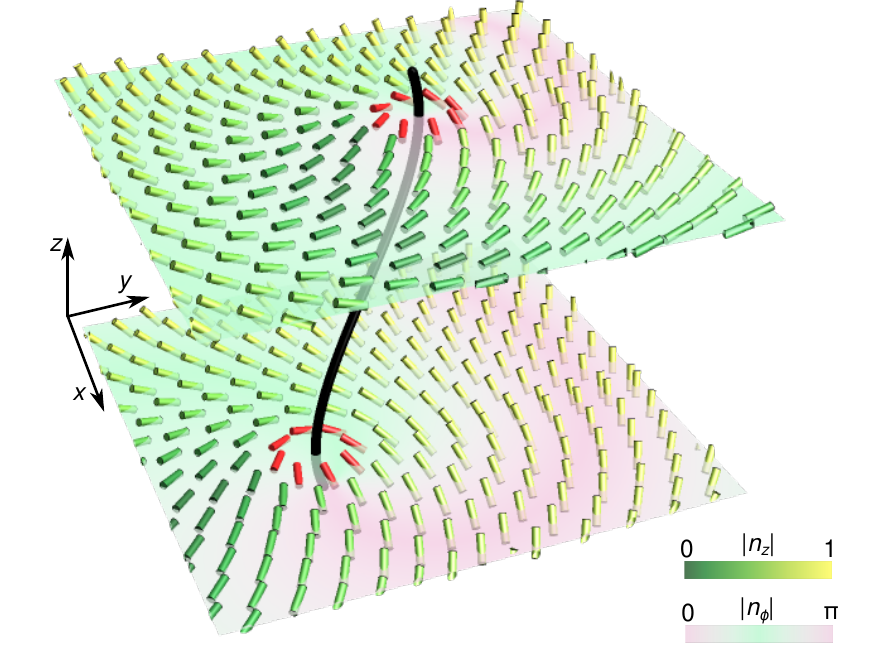}
\caption{{\bf Nematic bits (nbits)  pinned to a liquid crystal defect line.}
The local nematic director field $\vec{n}(\vec{r})$, indicated by cylindrical bars,  rotates by $\pi$ along closed curves encircling the defect line (black). The director field is colored by its out-of-plane component, $n_z(\vec{r})$, while $xy$-planes are colored by the director's azimuthal  orientation $n_\phi(\vec{r})$ relative to the $x$-axis. The near-field director profile (red) close to the defect line defines the nbit state (see also SI Fig.~S1). The vertical direction may be interpreted as either a spatial or a time dimension.   
}
\label{fig1}
\end{figure}

\begin{figure*}[t]
\centering
\includegraphics[width=\textwidth]{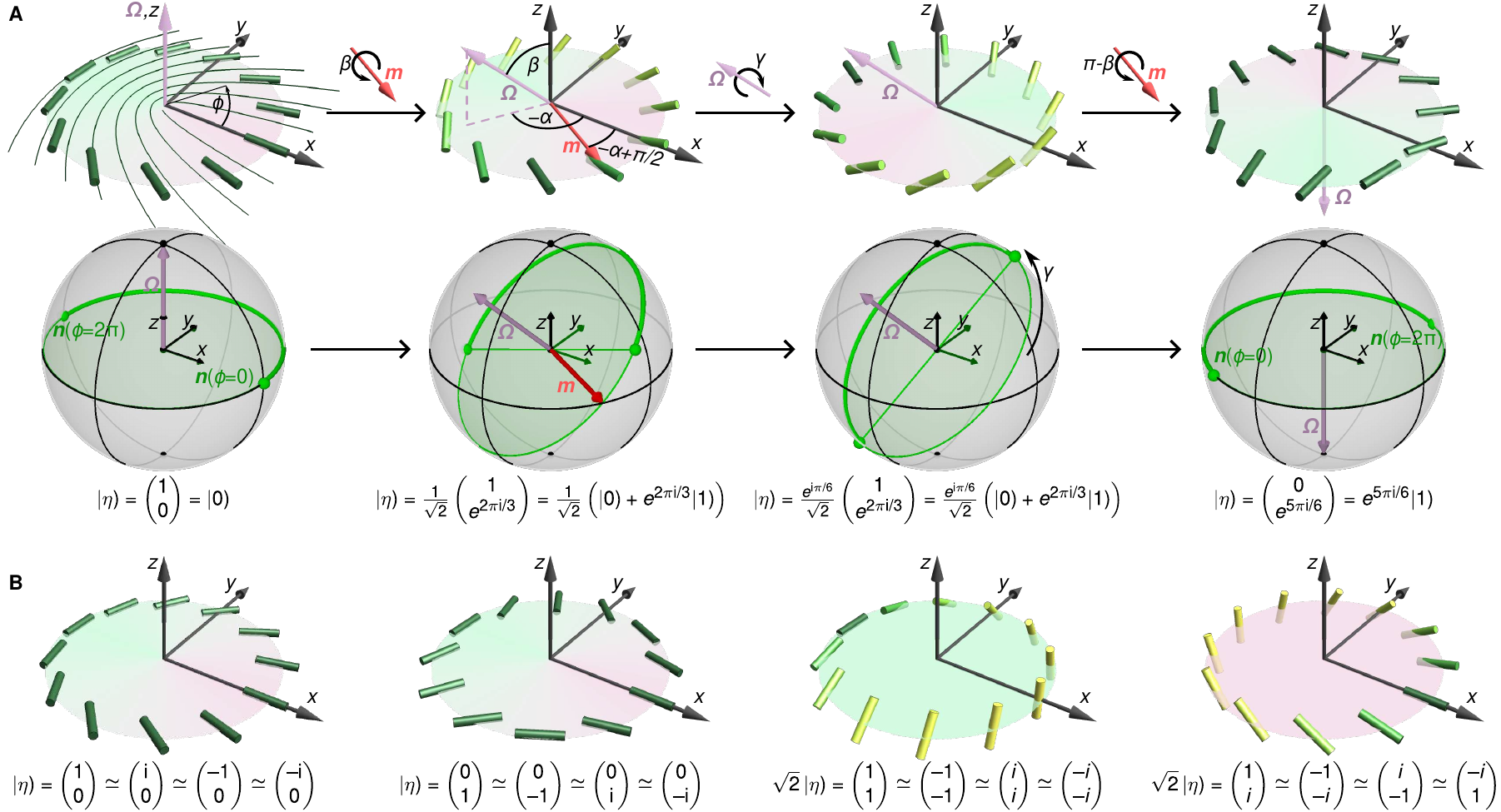}
\caption{{\bf Single-nbit transformations.}
(A) The director profile $\vec{n}_0(\vec{r})$, describing a $+1/2$-defect aligned with the positive $x$-direction, defines the nbit state~$|0)$. Arbitrary single-nbit states $|\eta)$ are obtained from~$|0)$ through simultaneous local director rotations, corresponding to near-field profiles $\vec{n}(\vec{r})=\eta \vec{n}_0(\vec{r})\eta^\dagger$ with $\eta$ given by~\eqref{eq:eta}; see also Movies 1 and 2. Each nbit~$|\eta)$ has a distinguished normal axis $\lambda \vec{\Omega}$, $\lambda \in \mathbb{R}$, spanned by an axial unit-vector $\vec{\Omega}$ perpendicular to the director field, $\vec{\Omega}\cdot\vec{n}(\vec{r})=0$; we choose $\vec{\Omega}$ to point along the positive $z$-direction for  $|0)$. 
Top: Example sequence showing how nbit $|0)$ can be continuously transformed to the nbit $e^{i 5\pi/6}|1)$ corresponding to a $-1/2$-defect profile. To this end,  $|0)$ is first homogeneously rotated around the axis $\vec{m}=(\cos(\alpha+\pi/2),\sin(\alpha+\pi/2),0)$ by an angle $\beta$, yielding an intermediate nbit having  a rotated normal $\vec{\Omega}$ with spherical polar coordinates $(\alpha,\beta)$. Subsequent director rotation around $\vec{\Omega}=(\alpha,\beta)$ by an angle $\gamma$ results in the arbitrary single-nbit state $|\eta)=|\alpha,\beta,\gamma)$. In the depicted example sequence, the second intermediate state is \mbox{$|\eta)=|-2\pi/3, \pi /4, \pi /3)$}.  Performing an additional rotation around $\vec{m}$ by $\pi-\beta$ leads to the $-1/2$-defect  nbit  $e^{i 5\pi/6}|1)$; for comparison,  the nbit $|1)$  corresponds to a $-1/2$-defect aligned with the $x$-axis.
Bottom:
On the nematic order-parameter unit sphere, an nbit director field $\vec{n}$ traces out a great semi-circle (thick green curves) from $\vec{n}(\phi=0)$ to $\vec{n}(\phi=2\pi)$, which lies  in a plane perpendicular to $\vec{\Omega}$. 
(B)  Four distinct nbits with $\vec{\Omega}$ pointing in the direction of $z$, $-z$, $x$, and $-y$, respectively. The symbol $\simeq$ indicates nbit states with identical near-field director profile but different global topology (see SI Fig.~S2).
}
\label{fig2}
\end{figure*}

\par

A widely studied class~\cite{1972deGennes,1988Lubensky_PRA,Bowick:2009aa} of soft matter systems which can be accurately controlled experimentally~\cite{2006Musevic_Science,2019Tai_Science} but whose computational potential has not yet been systematically investigated are nematic LCs. Composed of rod-like molecules, LCs can host  topological defects that are structurally robust against external perturbations, yet can be precisely manipulated through boundary conditions~\cite{2006Musevic_Science} and electric fields~\cite{2019Tai_Science}, as well as locally reconfigured with lasers~\cite{2011Tkalec_Science}. Building on recent theoretical work~\cite{CoparS_ProcRSocA469_2013} which identified a direct relation between string-like LC defects and quaternions,  we demonstrate here that such topological defects can be used as both classical binary and non-classical continuous nematic bits (nbits; Fig.~\ref{fig1}).  By deriving a reduced dynamical description from electro-nematic LC theory~\cite{AvelinoPP_SoftMatter7_2011,FukudaJ_PhysRevE81_2010}, we show how individual nbits, which correspond to points on the Poincar\'e-Bloch sphere~\cite{ToninelliE_AdvOptPhoton11_2019},  can be transformed in analogy with Pauli, Hadamard and other typical single-qubit gates~\cite{Ekert:2001aa} using electric fields. Generalizing to multi-nbit states,  we find that  nematoelastic interactions can cause strong correlations in an ensemble of nbit pairs, suggesting that such interactions can be used to realize logic functions. We confirm this prediction by demonstrating  universal classical NAND and NOR gates as well as generalized continuous logic functions in simulations for experimentally feasible nematic LC parameters. Our numerical results,  combined with quantitative estimates of the typical energetic costs and time scales associated with physical nbit-manipulations, suggest that nbit-circuits can be implemented with existing LC technology.

\section{Results}

\subsection*{Defining and transforming single nbit states}

\begin{figure*}[t]
\includegraphics[width=\textwidth]{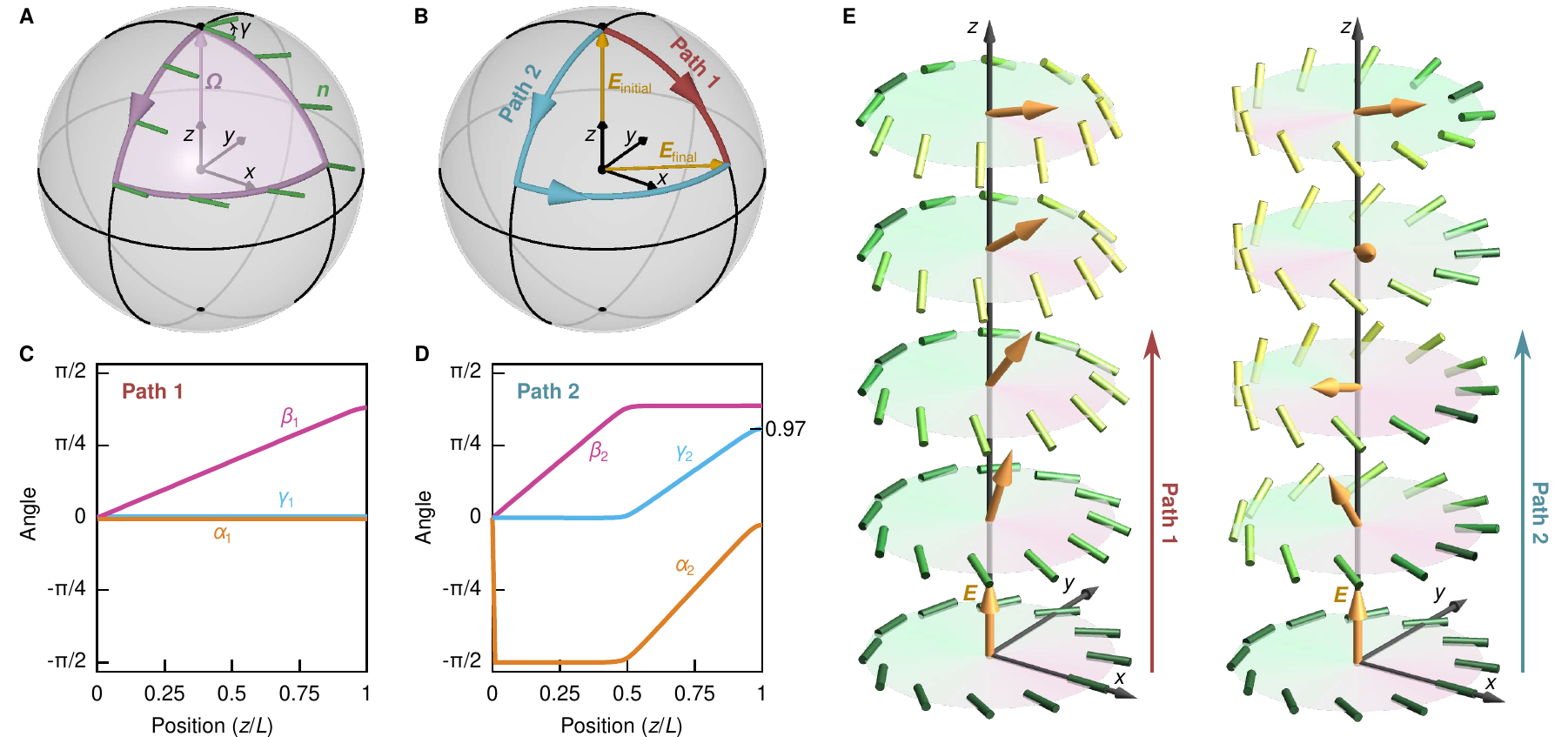}
\caption{{\bf Transforming nbits with electric fields.}
An initial nbit state in the plane $z=0$ can be transformed into an arbitrary target nbit at $z=L$ by applying a suitable electric field protocol $\vec{E}(z)$ that adjusts the orientation of the director normal vector $\vec{\Omega}$ along $z$ and realizes the global phase angle $\gamma$ in~\eqref{eq:eta_vector} by choosing an appropriate electric field path.
(A) The nbit phase $\gamma$  corresponds to a Berry phase and can be changed by moving $\vec{\Omega}$ around a closed path. By parallel transport, the change in  $\gamma$ is equal to the area enclosed by the path. 
(B) Two example  protocols  $\vec{E}(z)$, corresponding to distinct Paths 1 and 2, used in simulations of~\eqref{eq:nbit_dynamics}. Both paths go from the same initial field $\vec{E}_\text{initial}(z=0)$ to the same final field $\vec{E}_\text{final}(z=L)$, but result in a different final phase~$\gamma$.  
(C--D) Angles $\alpha$, $\beta$, $\gamma$ measured along $z$ in simulations of ~\eqref{eq:nbit_dynamics} for the two protocols $\vec{E}(z)$  from (B), starting with reference nbit $|0)$ at $z=0$.  (C) For Path~1, $\alpha_1$ rapidly jumps to its stationary value, while $\beta_1$ grows linearly with $z$  and~$\gamma_1$ remains constant. 
(D) For Path~2, $\alpha_2$ and $\beta_2$ approach similar final values as for Path 1, but $\gamma_2$ reaches a substantially larger final value close to the area of $\pi/3$ enclosed  by both paths.
(E) Simulated director configurations for the two electric field protocols from (B--D). Both protocols lead to final nbit states having the same normal vector $\vec{\Omega}$, which is determined only by the final electric field orientation at $z=L$. However, the final director profiles differ through a $\gamma_2$-rotation around $\vec{\Omega}$. Any final nbit state $|\eta)$ can be realized using this strategy.
}
\label{fig3}
\end{figure*}
Nematic LCs are assemblies of aligned rod-like molecules. By imposing  suitable boundary anchoring conditions~\cite{2006Musevic_Science,2011Tkalec_Science} or external electric fields~\cite{2019Tai_Science}, the molecules' global orientational order can be locally broken at singular defect points in 2D or lines in 3D~\cite{2012Alexander_RMP,2016MachonAlexander_PRX} (Figs.~\ref{fig1} and S1). Such topological defects~\cite{Mermin1979,Bowick:2009aa} present particularly promising candidates for robust information storage as their existence is protected
against thermal and other perturbations.  To define an nbit, we consider an elementary $+1/2$ defect in the $(z=0)$-plane of a 3D nematic LC, as shown in Fig.~2A top/left. In this configuration,  the defect  center-line passes through the coordinate origin,  all the molecule directors lie on the same plane (with normal $\vec{\Omega}$ along the $z$-axis), and the symmetry axis of the texture is aligned with the $x$-axis. We refer to this configuration as reference nbit state $|0)$ throughout and denote the associated director profile as  $\vec{n}_0(\vec{r})$, where $\vec{r}$ is the in-plane distance vector from defect line (Fig.~S1). Keeping the defect location fixed, arbitrary nbit states $|\eta)$ correspond to rotated director profiles $\vec{n}(\vec r)$,  obtained from $|0)$ by turning each director $\vec{n}_0(\vec{r})$ individually by the same angle $\theta$ about a common axis spanned by the unit vector $\vec{a}=(a_x,a_y,a_z)$. To highlight the mathematical parallels between nbits and qubits, we express these local director rotations in terms of quaternions~\cite{CoparS_ProcRSocA469_2013,BinyshJ_PhysRevLett124_2020}
\begin{equation}
\eta= \mathbb{1} \cos\frac{\theta}{2} + \mathrm{i} \left( a_x\sigma_x+a_y\sigma_y+a_z\sigma_z\right) \sin\frac{\theta}{2}
\label{eq:eta}
\end{equation}
where $\sigma_{x}, \sigma_y$ and $\sigma_z$ correspond to the Pauli matrices. The director profiles of the nbits $|\eta)$ and $|0)$ are then linked by quaternion product 
$
\vec{n}(\vec{r})=\eta\, \vec{n}_0(\vec{r})\,\eta^\dagger.
$ 
Since each quaternion of the form~\eqref{eq:eta} has an SU(2) matrix representation $\begin{pmatrix} c_1 & -\bar{c}_2\\c_2 & \bar{c}_1\end{pmatrix}$ with $|c_1|^2+|c_2|^2=1$, we can identify $|\eta)$ with the first column of the SU(2) matrix corresponding to $\eta$. 

\begin{figure*}[ht!]
\centering
\includegraphics[width=\textwidth]{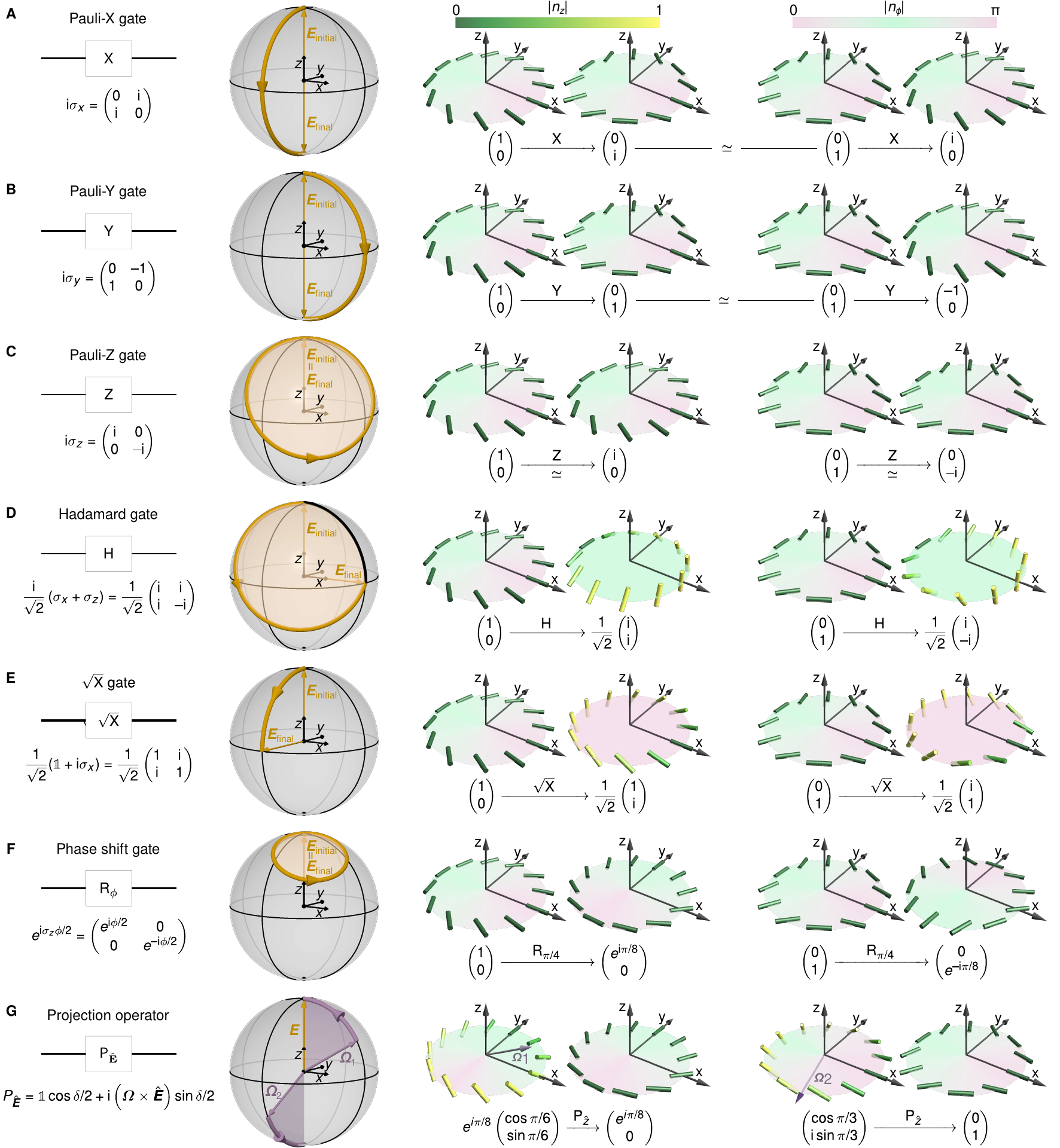}
\caption{{\bf Single-nbit logic gates and projection operator.}
Typical quantum logic gates (first column) acting on single-nbit states can be realized by rotating an applied electric field. The second column shows the electric field path required to realize logic operations on the $|0)$ bit. Columns 3 and 4 depict initial and final director configurations for logic gates applied to $|0)$ or $|1)$ nbits, respectively. 
(A--C)
Nematic Pauli gates perform a rotation by $\pi$ around the $x$, $y$ or $z$ axis, respectively. Pauli-X and Pauli-Y gates transform $|0)$ and $|1)$ into each other (up to a global phase). The Pauli-Z gate adds a phase $\pi/2$ to $|0)$ and a phase $-\pi/2$ to $|1)$.
(D)
The Hadamard gate rotates an nbit by $\pi$ around the $\hat{x}+\hat{z}$ axis, transforming $|0)$ and $|1)$ into tangential twist defect profiles.
(E)
The $\sqrt{\text{X}}$ gate performs a $\pi/2$ rotation around the $x$ axis, transforming $|0)$ and $|1)$ into radial twist defect profiles.
(F)~The phase shift gate  $\sqrt{\text{R}_\phi}$ induces a rotation by $\phi$ around the $z$ axis, preserving the angle $\beta$ of an nbit.  The action of $\sqrt{\text{R}_\phi}$ on the director fields of $|0)$ and $|1)$ is equivalent to a rigid rotation around the $z$ axis by $2\phi$ and $2\phi/3$, respectively.
(G) 
$\text{P}_{\vec{\hat{E}}}$ projects the director normal $\vec{\Omega}$ onto the axis of a sufficiently strong (SI)  electric field $\vec{\hat{E}}$ by rotating the nbit around $\vec{\Omega}\times\vec{E}$, where the rotation angle $\delta=\text{sign}(\vec{\Omega}\cdot\hat{\vec{E}})\text{acos}(\text{sign}(\vec{\Omega}\cdot\hat{\vec{E}})\vec{\Omega}\cdot\hat{\vec{E}})$ corresponds to the angle between $\vec{\Omega}$ and the electric field axis. For a projective measurement along the $z$-direction, nbits with $\vec{\Omega}$ pointing into the northern or southern hemispheres become projected onto the $|0)$ or $|1)$ subspaces, respectively. 
}
\label{fig4}
\end{figure*}

\par
To obtain a geometrically intuitive, equivalent representation of the spinor~$|\eta)$, it is convenient to decompose rotations of the reference nbit~$|0)$ into two steps (Fig.~\ref{fig2}A): First, the director field $\vec{n}_0$ of $|0)$ is rotated around the axis $\vec{m}=(\cos(\alpha+\pi/2),\sin(\alpha+\pi/2),0)$ by an angle $\beta$, yielding an intermediate nbit having a rotated normal $\vec{\Omega}$ with spherical polar coordinates $(\alpha,\beta)$; thereafter, the directors  are rotated around the new normal $\vec{\Omega}$ by an angle $\gamma$.  This gives the following geometric representation of an nbit
\begin{equation}
|\eta)=|\alpha,\beta, \gamma)=
e^{i\frac{\gamma}{2}}\begin{pmatrix}
\cos\frac{\beta}{2}\\
e^{-i\alpha}\sin\frac{\beta}{2}
\end{pmatrix},\label{eq:eta_spinor}
\end{equation}
where the global phase $\gamma$ corresponds to a Berry phase (Fig.~\ref{fig3}A).
As evident from~\eqref{eq:eta_spinor}, nbits for which $\gamma$ differs by $\pi$, $2\pi$ or $3\pi$ exhibit the same local director profile in the vicinity of the defect line (Fig.~\ref{fig2}B).  However, such locally  equivalent configurations can be distinguished by their global topology, as demonstrated in Movies 1 and 2 and Fig.~S2.
\par
Moreover, \eqref{eq:eta_spinor} makes explicit that the two basis states 
\begin{equation}
|0)=\begin{pmatrix}
1\\
0
\end{pmatrix},\qquad
|1)=\begin{pmatrix}
0\\
1
\end{pmatrix}
\end{equation}
correspond to $x$-axis aligned $+1/2$ and $-1/2$ defects, respectively (Fig.~\ref{fig2}B). Thus, an arbitrary nbit $|\eta)$ can be expressed as a normalized complex  linear combination of these two elementary defect  states:
\begin{equation}
|\eta)=
c_1|0)+c_2|1)
=
e^{i\frac{\gamma}{2}}
\left[
\cos\frac{\beta}{2}|0)
+
e^{-i\alpha}\sin\frac{\beta}{2}
|1)
\right],
\label{eq:eta_vector}
\end{equation}
with the complex vector superposition encoding physical director rotations. In particular, the above construction shows that single-nbit states are mathematically equivalent to single-qubit states. It should be stressed, however, that despite sharing a mathematically equivalent state-space, nbits and qubits are not physically equivalent.  In particular, classical nbits are governed by a dissipative dynamics, whereas qubits obey wave-like Schr\"odinger-Pauli dynamics. As consequence, multi-qubit states can produce interference phenomena that are unlikely to find  direct counterparts in nbit-systems, implying differences in their computational capabilities. Notwithstanding such differences, we will see below that nbits can be used to implement both classical binary logic functions as well as generalized continuous logic functions.

\subsection*{Controlling nbits with electric fields}

The ability to access and manipulate individual nbits is essential for the experimental implementation of nematic LC computers.
To demonstrate a practically feasible protocol, we generalize recent advances in the electro-nematoelastic transformation of LC defects~\cite{AvelinoPP_SoftMatter7_2011,FukudaJ_PhysRevE81_2010}. Starting from the field equations of liquid crystal hydrodynamics, one can show~(SI) that the adiabatic relaxation dynamics of an nbit-quarternion~\eqref{eq:eta} in a slowly varying, sufficiently strong electric (SI) field  $\vec{E}=E_x\sigma_x+E_y\sigma_y+E_z\sigma_z$ is determined by the matrix evolution equation
\begin{equation}
\frac{\mathrm{d} \eta}{\mathrm{d} t}=\frac{K}{\Gamma}\frac{\mathrm{d}^2 \eta}{\mathrm{d} z^2}
-   \frac{\epsilon_\text{a}}{4 \Gamma} \mathrm{Tr}\left(\eta\sigma_z \eta^\dagger \vec{E}\right) \vec{E}\eta\sigma_z   -\lambda \eta,
\label{eq:nbit_dynamics}
\end{equation}
where $K$ is the elastic constant of the nematic director field, $\Gamma$ denotes the rotational diffusion constant, and $\epsilon_\text{a}$ is the dielectric anisotropy. The Lagrange multiplier $\lambda$ preserves the SU(2) structure of $\eta$. In the remainder, we shall focus on LCs with negative dielectric anisotropy, $\epsilon_\text{a}<0$, such as N-(4-Methoxybenzylidene)-4-butylaniline (MBBA)~\cite{KlemanM};  analogous control strategies can be devised for materials with $\epsilon_\text{a}>0$.
\par
For LCs with $\epsilon_\text{a}<0$,~\eqref{eq:nbit_dynamics} implies that the director normal axis $\vec{\Omega}$ of an nbit  prefers to align with the electric field~$\vec{E}$. This means that one can program the polar coordinates $(\alpha,\beta)$ of $|\eta)$ through the instantaneous direction of $\vec{E}$. Additionally, the global Berry phase $\gamma$ of an nbit can be set by moving  $\vec{\Omega}$ around a closed path (Fig.~\ref{fig3}A). Suitably designed spatial (or temporal) electric field protocols can thus be used to transform an initial nbit into any desired target nbit (Fig.~\ref{fig3}).  Figures~\ref{fig3}B--E show two example protocols $\vec{E}(z)$  that transform nbit $|0)$, which is localized in the plane $z=0$, into distinct target nbits at $z=L$. The two protocols connect the same initial and final field values, $\vec{E}(z=0)$  and $\vec{E}(z=L)$, via two different paths 1 and 2  (Fig.~\ref{fig3}B), so that the final nbit states have similar normal coordinates $(\alpha,\beta)$  but differ in their global phase angles $\gamma$ (Fig.~\ref{fig3}C,D). The corresponding nbit director fields,  computed numerically (Methods) from the stationary solutions of~\eqref{eq:nbit_dynamics} along both paths, can be seen in Fig.~\ref{fig3}E. Importantly, the same electric control strategy can be used to implement nematic logic gates.

\subsection*{Nematic logic gates and projective measurements}

Since, according to~\eqref{eq:eta_vector}, single-nbit states $|\eta)$ are equivalent to elementary qubits, one can implement direct nematic analogues of many standard quantum logic gates~\cite{NielsenM} by applying suitable electric field profiles. Figure~\ref{fig4} summarizes realizations of commonly used gates, showing the electric field protocols required to transform $|0)$ and also the final nbit states when logic operations are performed on $|0)$ or $|1)$, respectively. Specifically, we demonstrate the actions of the three nematic Pauli gates (Fig.~\ref{fig4}A--C), the Hadamard gate (Fig.~\ref{fig4}D), $\sqrt{\text{X}}$ gate (Fig.~\ref{fig4}E) and the phase shift gate (Fig.~\ref{fig4}F). As a general practical rule, the initial electric field of the gate needs to be aligned with the normal axis $\vec{\Omega}$ of the nbit $|\eta)$ to which the gate operation is applied. Experimentally, the determination of $\vec{\Omega}$ can be done optically and non-invasively in LC materials~\cite{PosnjakG_SciRep6_2016}, in contrast to quantum systems. Another interesting class of operations,  corresponding to projective measurements,  can be realized by applying electric fields with a fixed direction. For example, by fixing the field vector $\vec{E}$  in~\eqref{eq:nbit_dynamics} parallel to the $z$-direction, all nbits with $\vec{\Omega}$ pointing into the norther hemisphere become projected onto the $|0)$ subspace, whereas all nbits with $\vec{\Omega}$ in the southern hemisphere become projected onto the $|1)$ subspace (Fig.~\ref{fig4}G). These single-nbit operations provide a basis for implementing computations in nematic LC systems.

\begin{figure*}[t!]
\centering
\includegraphics[width=\textwidth]{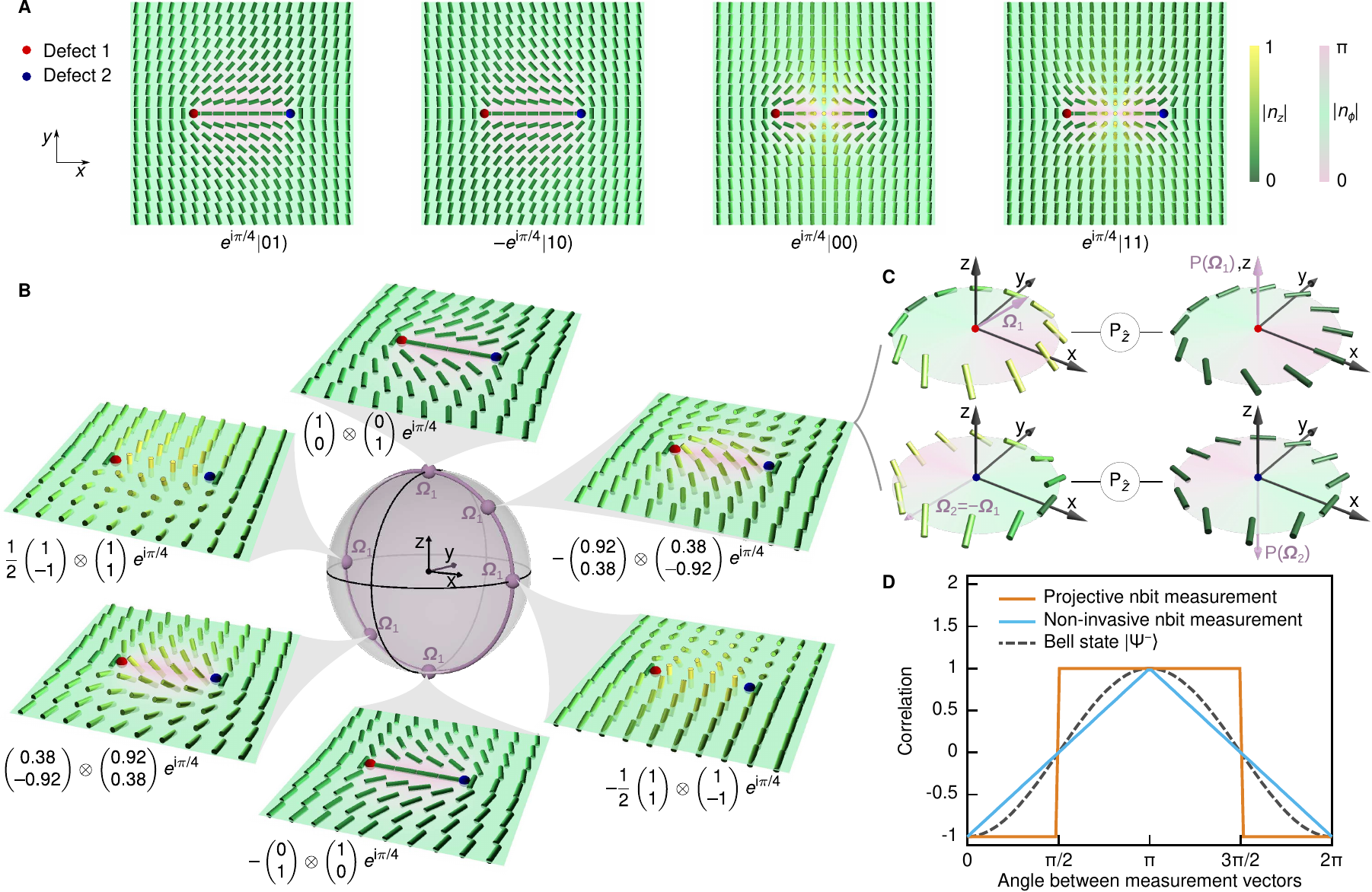}
\caption{{\bf Two-nbit states and Bell-type correlations.} 
A two-nbit product state $|\eta_1\eta_2)=|\eta_1)\otimes|\eta_2)$ is a topological defect pair, with the local director field structure around the defects 1 and 2 defining the single-nbits $|\eta_1)$ and $|\eta_2)$, respectively. 
(A) The two-nbit states $e^{i\pi/4}|01)$ and $-e^{i\pi/4}|10)$ are free-energy minima, whereas $e^{i\pi/4}|00)$ and $e^{i\pi/4}|11)$ exhibit an umbilic soliton~\cite{2016MachonAlexander_PRX} at the center. 
(B)
Representative examples of energetically equivalent (Fig.~S5) two-nbit product states with the director far-field parallel to the $y$ direction, selected from the ensemble manifold $\Psi^-$ of anti-parallel nbit states  with $\vec{\Omega}_1=-\vec{\Omega}_2$. The two-nbit states in the $\Psi^-$ ensemble can be strongly correlated under projective measurements due to nematoelastic interactions.
(C)
A global projective measurement, performed by applying a $z$-aligned electric field, $\vec{\hat{E}} || \vec{\hat{z}}$, that projects both $\vec{\Omega}_1$ and $\vec{\Omega}_2$ along the $z$ direction (Fig.~4G), collapses each two-nbit state in the $\Psi^-$ ensemble to a $\pm 1/2$-defect pair.
(D)
The  choice of the local measurement procedure determines the observed nbit correlations, defined as the ensemble average $\langle \text{sign}(\vec{\Omega}_1\cdot\vec{\hat{v}}_1)\cdot\text{sign}(\vec{\Omega}_2\cdot\vec{\hat{v}}_2)\rangle_{\Psi^{-}}$ where  $\vec{\hat{v}}_1$ and $\vec{\hat{v}}_2$ are the local measurement vectors applied to nbit 1 and 2, respectively. Non-invasive measurements produce the classically expected piecewise linear correlation dependence on the relative angle between $\vec{\hat{v}}_1$ and $\vec{\hat{v}}_2$  (blue).  By contrast, projective measurements performed by successively  applying local electric fields with orientations $\vec{\hat{E}}_1=\vec{\hat{v}}_1$ and  $\vec{\hat{E}}_2=\vec{\hat{v}}_2$ to the nbits can result in nearly perfect correlation or anticorrelation (orange), provided the time delay between the two measurements is sufficiently long to allow for nematoelastic relaxation.
}
\label{fig5}
\end{figure*}

\subsection*{Two-nbit states}

The above framework can be generalized to systems with two and more interacting nbits.  To construct their mathematical description, we consider  configurations of two nearby defects in the $xy$-plane (Fig.~\ref{fig5}A and Figs.~S3, S4). The defects represent nbits $|\eta_1)$ and $|\eta_2)$, respectively, and we can describe their joint state  by the tensor product $|\eta_1)\otimes |\eta_2)$. Analogous to the single-nbit case above,  a $(+1/2,-1/2)$ defect pair aligned with the $x$-axis defines the reference state $e^{i\pi/4}|0)\otimes|1)\equiv e^{i\pi/4}|01)$, where the global phase factor reflects the rotation of the $-1/2$ defect (Fig.~\ref{fig5}A, left).  Arbitrary two-nbit product states can be realized by rotating the local director fields around each defect, with the individual phases $\gamma_1$ and $\gamma_2$ determining the global two-nbit phase $\gamma=\gamma_1+\gamma_2$. For example, by rotating the directors of each of the two nbits in the reference state $e^{i\pi/4}|0)\otimes|1)$ by an angle $\pi$ around the $y$-axis, corresponding the action of the Pauli-Y gate (Fig.~\ref{fig4}B), one obtains the state \mbox{$-e^{i\pi/4}|1)\otimes|0)\equiv -e^{i\pi/4}|10)$} shown in the second panel of  Fig.~\ref{fig5}A.  
General nematic product states are given by $e^{i\gamma_{ij}}|i) \otimes |j)$, corresponding to pairs of suitably rotated topological defects (Fig.~\ref{fig5}A,B). Ensembles of two-nbit states can exhibit  strong statistical correlations, facilitated by nematoelastic interactions of the director fields.

\subsection*{Nematoelastic interactions}

Nematoelastic  interactions may provide a resource for nematic computation. For example, when the defects 1 and 2 forming a  two-nbit state are embedded into a homogeneous director far-field, their energetic equilibrium configurations correspond to director profiles in which the local normal vectors $\vec{\Omega}_1$ and $\vec{\Omega}_2$ are perfectly antialigned, $\vec{\Omega}_1=-\vec{\Omega}_2$ (Fig.~\ref{fig5}B). Moreover, the common axis of $\vec{\Omega}_1$ and $\vec{\Omega}_2$ can point in an arbitrary direction, defining an ensemble $\Psi^-$ of energetically degenerate states if the nematic material can be appropriately described by a single dominant elastic constant~$K$ (Fig.~\ref{fig5}B). This leads to a strong nematoelastic coupling:  if one nbit is slowly rotated, the other nbit will adiabatically follow to ensure   that $\vec{\Omega}_1=-\vec{\Omega}_2$. Similarly, by enforcing a umbilic soliton~\cite{2016MachonAlexander_PRX} in the system~(Fig.~\ref{fig5}A), one can realize an ensemble $\Phi^+$ of two-nbit states with $\vec{\Omega}_1=\vec{\Omega}_2$ (Fig.~S6). The nonlocal nematoelastic coupling of the nbit orientations can lead to Bell-type correlations when projective measurements are performed on individual nbits (Fig.~\ref{fig5}C,D). From a practical perspective,  this suggests that nematoelastic interactions can be exploited as a computational resource.

\subsection*{Multi-nbit logic operations}

\begin{figure*}[t!]
\centering
\includegraphics[width=\textwidth]{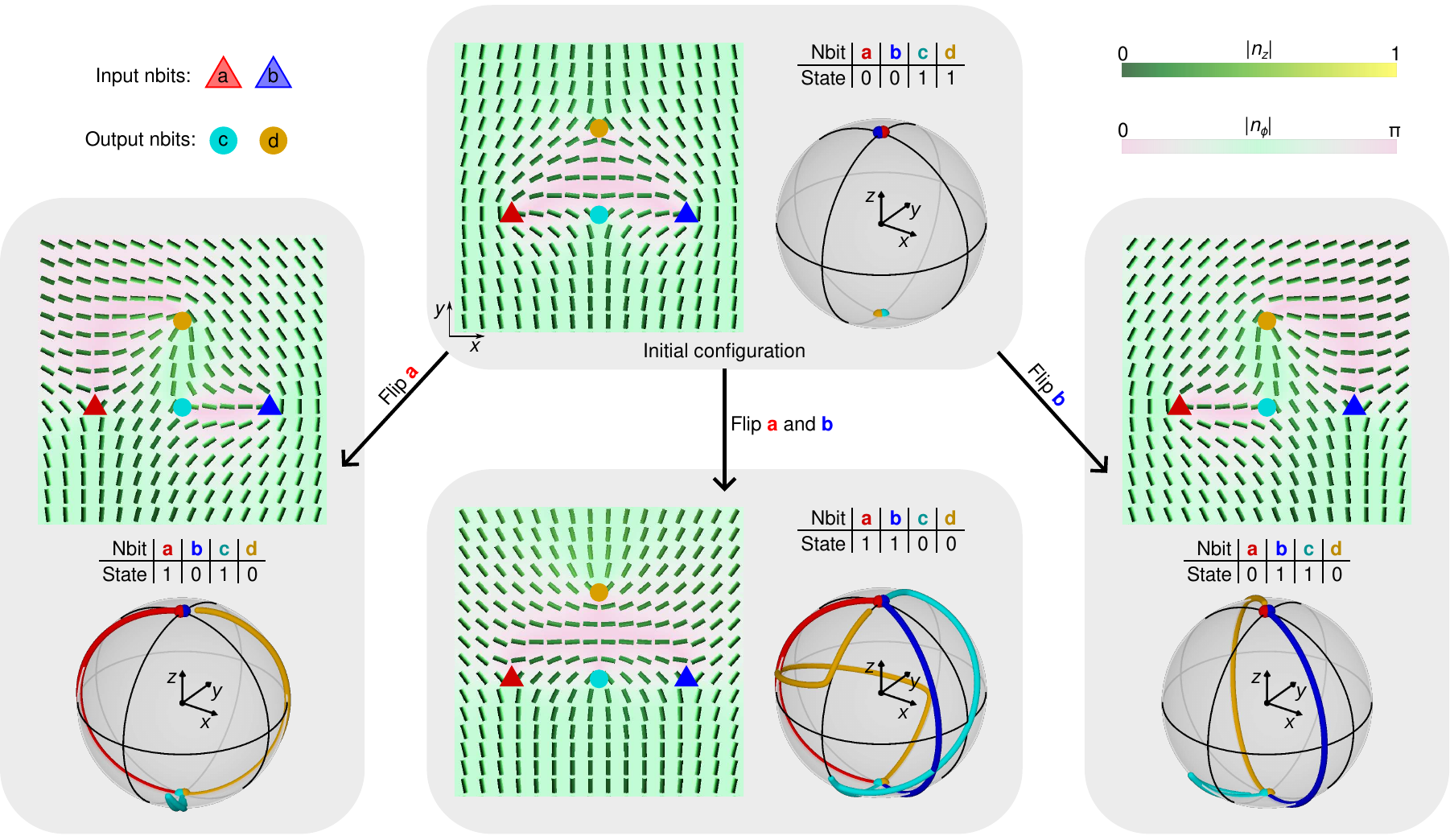}
\caption{
{\bf Universal classical logic gates.}
In the initial configuration (top panel) the two input nbits `a' and `b' are in the 0 state (corresponding to $|0)$) and two output nbits  `c' and `d'  in the 1 state (corresponding to $|1)$). The initial configuration is transformed by changing the director field in the vicinity of the input nbits and observing the nematoelastic response of the output nbits; the Poincar\'e-Bloch spheres show the evolution of the nbit states during each reconfiguration process.  Left panel: nbit `a' is flipped into the 1 state by rotating its local director field around the $(\cos(-3\pi/8),\sin(-3\pi/8),0)$ axis, causing the output nbit `d' to change from 1 to 0  while leaving `b' and `c' unchanged (up to a phase factor). Right panel: the director of input nbit `b' is rotated around the $(\cos(-\pi/8),\sin(-\pi/8),0)$ axis, causing a flip in the state of the output nbit `d'. Bottom-middle  panel: Simultaneous flipping of the two input nbits `a' and `b' results in a 0 state for both output nbits. Thus, nbits \lq c\rq{} and \lq d\rq{} obey classical NAND and NOR  gate transformation rules, respectively.
}
\label{fig6}
\end{figure*}

Utilizing strong correlations between nbits, we can perform classical logic operations in systems of multiple qubits. Figure~\ref{fig6} shows a logical operation on a system of four nbits, where in the initial configuration, two input nbits  \lq a\rq{} and \lq b\rq{} are in a $|0)$ state (up to a phase factor), and two output nbits \lq c\rq{} and \lq d\rq{} are in a $|1)$ state. One or both input nbits are then flipped into the $|1)$ state, tracking the output response. After the equilibrium is reached, output nbits are observed to be in either the $|0)$ or $|1)$ state. The combined truth table reveals that the nbit transformations in Fig.~\ref{fig6} realize a universal classical NAND gate for output nbit~\lq c\rq  and a universal NOR gate for output nbit \lq d\rq.

\begin{figure*}[t!]
\centering
\includegraphics[width=\textwidth]{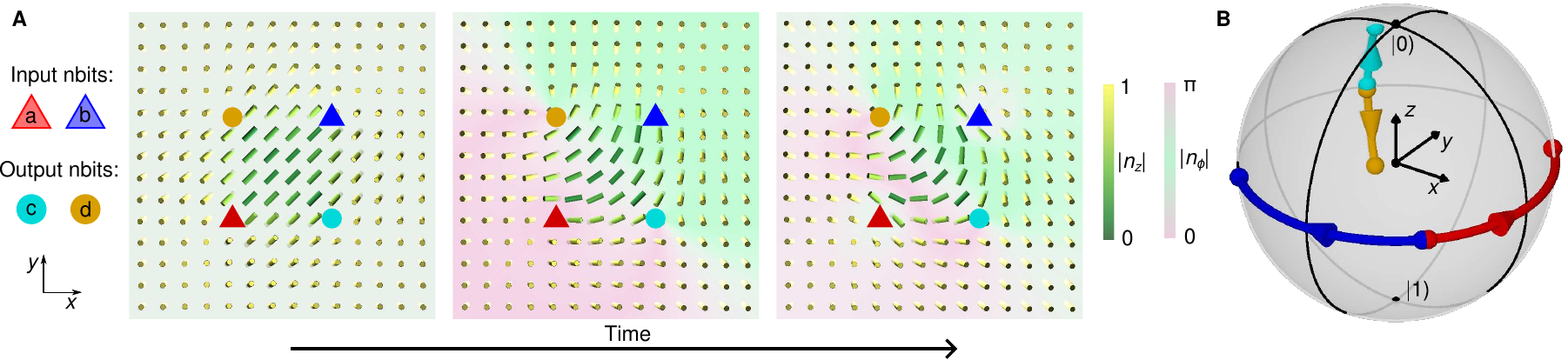}
\caption{
{\bf Multi-nbit operations beyond classical logic functions.}
By varying the defect positions and initial multi-nbit configuration, it is possible to implement generalized continuous  logical operations on a Poincar\'e-Bloch sphere, where input and output nbits are not limited to switching between $|0)$ and $|1)$, but can assume any superposition of these states. The figure shows an example, in which defects are arranged in a square pattern and input nbits `a' and `b' are transformed from $|a)_\text{init},|b)_\text{init}\propto  |0) + e^{\mathrm{i}\pi/4}|1) $ to $|a)\propto |0) + e^{-\mathrm{i}\pi/4} |1)$ and $|b)\propto |0) + e^{\mathrm{i}3\pi/4} |1)$, respectively (prefactors are ommited in the notation). Output nbits `c' and `d' respond to the input configuration, transforming from $|c)_\text{init},|d)_\text{init}\propto |0) + e^{-\mathrm{i}3\pi/4}|1) $ to $|c)\propto 0.87|0) + 0.50e^{-2.32\mathrm{i}}|1)$ and $|d)\propto 0.50|0) + 0.87e^{-2.32\mathrm{i}}|1)$, respectively.
(A) Timeline of input and output nbit transformations.
(B) Paths of nbit transformations traced on a Poincar\'e-Bloch sphere.
}
\label{fig7}
\end{figure*}

A general multi-nbit logic operation does not involve only discrete 0 and 1 bits, but is in principle determined by a mapping from Poincar\'e-Bloch spheres of the input states to the Poincar\'e-Bloch sphere of the output. We show an example of such a non-digital operation in Fig.~\ref{fig7}, where input and output nbits are always in a superposition of $|0)$ and $|1)$ states. For the chosen spatial arrangement and initial configuration of the nbits, transforming the input nbits along the equator of the Poincar\'e-Bloch sphere results in a change of the polar angle on the Poincar\'e-Bloch sphere of the output nbits.

\section{Discussion}

\subsection*{Non-invasive vs. projective nbit measurements}
State-of-the-art optical techniques~\cite{PosnjakG_SciRep6_2016} can measure the defect textures in LC materials without perturbing them, whereas applied electric fields can be used to reorient LCs~\cite{2019Tai_Science} along a preferred axis (Figs.~\ref{fig4}G and \ref{fig5}C). This means that nbit systems allow for both non-invasive and projective measurements. Depending on the details of the experimental measurement protocol and the characteristic time scale of the nematoelastic interactions, two-nbit ensembles can exhibit statistical correlations that are weaker or stronger than those of quantum Bell states. 
Such correlations can provide a computational resource for realizing logic operations in LCs.
We demonstrate the dependence of the correlations on the measurement protocol for the $\Psi^{-}$ ensemble, consisting of two-nbits states with  $\vec{\Omega}_2=-\vec{\Omega}_1$ where $\vec{\Omega}_1$ is uniformly distributed on the two-dimensional unit sphere (Fig.~\ref{fig5}B). We compare the nematic correlation strength with the expected spin correlations for the maximally entangled two-qubit Bell-state~\cite{1964Bell,NielsenM} \mbox{$
|\Psi^-\rangle=\left(|0\rangle\otimes |1\rangle - |1\rangle\otimes |0\rangle\right)/\sqrt{2}$} in a standard projective quantum measurement. If the state of the first qubit in $|\Psi^-\rangle$ is measured along some direction $\vec{\hat{v}}_1$ and that of the second qubit along $\vec{\hat{v}}_2$, then the observed correlations are known to be stronger than for optimized classical \lq local realist\rq~\cite{NielsenM} imitations (Fig.~\ref{fig5}D). Indeed, in the case of a non-invasive nbit measurement, the analytically calculated ensemble-averaged correlation function \mbox{$\langle \text{sign}(\vec{\Omega}_1\cdot\vec{\hat{v}}_1)\cdot\text{sign}(\vec{\Omega}_2\cdot\vec{\hat{v}}_2)\rangle_{\Psi^{-}}$} exhibits the classically expected linear dependence on the angle between $\vec{\hat{v}}_1$ and $\vec{\hat{v}}_2$ (blue curve in Fig.~\ref{fig5}D).  By contrast, projective nbit measurements can lead to substantially stronger correlations due to the nematoelastic coupling between the nbits. Applying a global electric field along the $z$-axis projects the nbits  of a $\Psi^-$ configuration onto a pair of $\pm 1/2$ defects, corresponding to states locally equivalent to $e^{i\pi/4}|01)$ or $-e^{i\pi/4}|10)$ (Fig.~\ref{fig5}C). Similarly, a global projection along  $x$ transforms $\Psi^-$ configurations into states that are proportional to tensor products of $|+)\equiv \left[|0)+|1)\right]/\sqrt{2}$ and $|-)\equiv \left[|0)-|1)\right]/\sqrt{2}$. Particularly interesting correlations due to nematoelastic interactions arise when two local projective measurements along different axes are performed successively on the two nbits. To see this, let us assume that nbit~1 of a $\Psi^-$ configuration is first projected along $\vec{\hat{v}}_1$ by a local electric field with orientation $\vec{\hat E}_1=\vec{\hat{v}}_1$.  After the measurement, $\vec{\Omega}_1$ will point along $+\vec{\hat{v}}_1$ or $-\vec{\hat{v}}_1$ and,  given sufficient time, the nematoelastic interactions will reorient the axis of the second nbit until $\vec{\Omega}_2=-\vec{\Omega}_1$ is reached.   Thus, subjecting the reoriented nbit 2 to a second projective measurement along $\vec{\hat E}_2=\vec{\hat{v}}_2$ yields the strongest possible correlation (orange curve in Fig.~\ref{fig5}D). 
In practice, one can tune the correlation strength in an nbit system by adjusting the time delay between the projection measurements relative to the nematoelastic relaxation time scale.

\subsection*{Time scales and energetic costs}
The rate at which logic gate operations on nbits can be performed depends on both the elastic relaxation time scale of the LC material, $\tau_\text{elastic}\simeq \Gamma R^2/K$,  where $R$ is the  characteristic cylindrical radius of the nbit line-defect domain, and the material's response time in an applied electric field, $\tau_\text{electric}\simeq {\Gamma}/(|\epsilon_a| E^2)$. Elementary nbit transformations require $\tau_\text{electric}\ll \tau_\text{elastic}$ to realize non-direct paths between the initial and final nbit direction (Fig.~\ref{fig3}A,B). This conditions is satisfied for commonly used LC materials; for example, considering MBBA~\cite{KlemanM,deGennes} with  $\Gamma=0.076\,\text{Pa s}$, $R=10\,\upmu\text{m}$, $K=5\,\text{pN}$, $\epsilon_\text{a}=-0.7\epsilon_0$ where $\epsilon_0$ is the vacuum permittivity, and $E=0.5\,\text{V/}\upmu\text{m}$,  one finds $\tau_\text{elastic}=1.5\,\text{s}$ and $\tau_\text{electric}=0.049\,\text{s}$. To ensure that the director field can adiabatically follow the electric field path, the rate at which the electric control field is changed must be slower than the electric response rate~$1/\tau_\text{electric}$. Furthermore, for two-nbit measurements, the elastic relaxation scale separates the regimes of fast (non-invasive) measurements ($\tau_\text{meas}\ll\tau_\text{elastic}$) and slow projective measurements ($\tau_\text{meas}\gg\tau_\text{elastic}$), see Fig.~\ref{fig5}D.

The energetic cost of manipulating an nbit can be estimated from the energy dissipation formula Eq.~(S4). For example, transforming nbit $|0)$ into $|1)$ over a time period~$\tau$ requires an energy of (SI)
\begin{equation}
E\simeq\frac{\pi^3\Gamma}{4\tau} L R^2,
\label{eq:diss}
\end{equation}
where $L$ is the length of the defect line segment.~\eqref{eq:diss} shows that fast reorientations require proportionally more energy.  By considering 
$E\sim k_B T$,  where $k_B$ is the Boltzmann constant and $T$ denotes temperature, we can estimate the time scale  $\tau_B$ over which an nbit becomes flipped by thermal fluctuations. Adopting MBBA parameters as above and assuming an nbit volume $LR^2\sim(10\,\mu\text{m})^3$ at room temperature, one finds $\tau_B=1.7\,\text{days}$, suggesting that nbits are thermally stable for typical operational time scales in the seconds range. \eqref{eq:diss} implies that $\tau_B$ can be increased by increasing the nbit-volume, at the expense of requiring stronger electric fields for logical operations.

\section{Conclusions}

Recent theoretical~\cite{2012Alexander_RMP,2016MachonAlexander_PRX,LongC_SoftMatter17_2021,TangX_PhysRevE101_2020,PearceDJG_SoftMatter17_2021} and experimental~\cite{2006Musevic_Science,2011Tkalec_Science,2011Kamien_Science,2019Tai_Science,EmersicT_ScienceAdvances5_2019} advances enable unprecedented control over topological structures in nematic LCs. Our above analysis provides a conceptual framework for storing and processing information in  the textures of nematic fluids which, unlike hardwired solid-state devices, can be reconfigured and adapted. Similar to the development of quantum computation over the last three decades~\cite{1997DeutschJozsa,1997Shor_SIAM}, the next challenge is to identify classes of problems that can be efficiently solved with algorithms that utilize the single-nbit and multi-nbit gate operations (Fig.~\ref{fig4} and~\ref{fig6}).  The mathematical parallels between nbit and qubit systems can offer helpful guidance in this context. In the single-nbit case, the identification of director profiles with SU(2) matrices led to~\eqref{eq:eta_vector}, which reflects the Hopf-fibration~\cite{MosseriR_JPhysAMathGen34_2001}  of the Poincar\'e-Bloch sphere and allowed us to interpret nbits as superpositions of $+1/2$ and $-1/2$ defects. Therefore, a particularly interesting question, both from a mathematical and practical perspective, is whether it is possible to construct nematic analogues of the $S^7$  Hopf fibration that encodes the entanglement of two-qubit states~\cite{MosseriR_JPhysAMathGen34_2001,2016MachonAlexander_PRX}. 
More generally, however,  the above results suggest that nbits can provide a fruitful new paradigm for exploring the computational potential of soft matter systems.

\section{Methods}

\begin{small}
\noindent{\bf Characterization of nbit defect profiles.} 
A starting point for creation of nbit state is a reference $|0)$ defect profile
\begin{equation}
\vec{n}_0=\left(\cos\phi/2, \sin\phi/2,0 \right),
\end{equation}
where $\phi$ is the azimuthal position on the $xy$-plane. $\vec{n}_0$ is transformed according to $\vec{n}(\vec{r})=\eta \vec{n}_0\eta^\dagger$ and~\eqref{eq:eta} into a general nbit configuration:
\begin{eqnarray}
\vec{n}(\phi)&=&-\sin\left(\frac{\phi}{2}+\gamma-\alpha\right)\vec{m}+
\notag\\
&&\notag
\cos\left(\frac{\phi}{2}+\gamma-\alpha\right)\left(\cos\beta\vec{t}\times\vec{m}+\sin\beta\vec{t}\right),\label{eq:spinor}
\end{eqnarray}
where $\vec{m}=(\cos(\alpha+\pi/2),\sin(\alpha+\pi/2),0)=\vec{n}\left(\phi=2\alpha-2\gamma+\pi\right)$ corresponds to the director orientation at a point where it lies in the $xy$-plane, and $\vec{t}$ is a defect line tangent that is in the paper always aligned~along~$z$-axis.
\\

\noindent{\bf Simulation methods.}
Stationary solutions of~\eqref{eq:nbit_dynamics} were found numerically by finite difference approach on a cubic space-time mesh. In Fig.~\ref{fig3},  a value $\epsilon_\text{a}|\vec{E}|^2L^2/(4K)=400$ is used and boundary conditions of $\eta=\mathbb{1}$ at $z=0$ and ${\mathrm{d}\eta}/{\mathrm{d}z}=0$ at $z=L$ are applied. 

Nematic fields in Fig.~\ref{fig5}A and B, Fig.~\ref{fig6} and  Fig.~\ref{fig7} were obtained by numerical minimization of the free energy  with density of
\begin{equation}
f=\frac{A}{2}Q_{ij}Q_{ji}+\frac{B}{3}Q_{ij}Q_{jk}Q_{ki}+\frac{C}{4}\left(Q_{ij}Q_{ji}\right)^2+\frac{L}{2}\frac{\partial Q_{ij}}{\partial x_k}\frac{\partial Q_{ij}}{\partial x_k},
\end{equation}
where $Q$ is the nematic tensor order parameter, $L$ is nematic tensorial elastic constant, and $x_k$ the $k$-th spatial coordinate. Adopting the same values of the  phase parameters $A,B,C$ as in Ref.~\cite{2011Tkalec_Science}, the free energy minimum is found using a gradient descent method on a rectangular grid. The defects were pinned by locally decreasing the degree of order. Only the region of the simulation box that contains defects is shown in the figures.
\\
\end{small}

\begin{small}
\noindent{\small\bf Acknowledgements}\\
\noindent 
The authors thank Jack Binysh and Simon \v{C}opar for help with defect visualizations, and  
Gareth P. Alexander, Simon \v{C}opar, Thomas Machon and Miha Ravnik for helpful discussions and comments.
This work was supported by the Slovenian Research Agency (ARRS) under contracts P1-0099 and N1-0124 (\v{Z}.K.)  and the Robert E. Collins Distinguished Scholarship Fund (J.D.).
\end{small}


\newpage

\onecolumngrid

\section{Appendix}

\appendix

\section{Nbit solutions for nematic defects}
Stationary solutions for the director field $\vec{n}$ of a nematic liquid crystal correspond to minima of the free energy with density~\cite{KlemanM}
\begin{equation}
f=\frac{K}{2}\left(\nabla\vec{n}\right)^2-\frac{\epsilon_\text{a}}{2}\left(\vec{n}\cdot\vec{E}\right)^2,\label{eq:free_energy}
\end{equation}
where $K$ denotes the elastic constant of the director field deformations, $\epsilon_\text{a}$  is the anisotropy of the dielectric tensor, and $\vec{E}$ is the electric field.
Out of equilibrium,  the director field $\vec{n}$ is driven towards a free energy minimum by a \quotes{molecular field} $\vec{h}=-{\delta f}/{\delta \vec{n}}$, yielding the relaxation dynamics 
\begin{equation}
\Gamma\dot{\vec{n}}=\vec{h}-\lambda\vec{n}=K\nabla^2\vec{n}-\epsilon_a\left(\vec{n}\cdot\vec{E}\right)\vec{E}-\lambda\vec{n},
\label{eq:field}
\end{equation}
where $\Gamma$ is the rotational diffusion constant (also known as $\gamma_1$ in the literature) and $\lambda$ is a Lagrange multiplier that preserves the normalization $|\vec{n}|=1$.
\par
The nematic director field can contain singularities in form of defect lines, often appearing in half-integer form, in which case the director field rotates by an angle of $\pi$ when circumnavigating a defect line. To obtain a director field solution close to a half-integer defect line, the Laplace operator is written in cylindrical coordinates. Then, in close proximity to the defect line, the molecular field reduces in leading order to
\begin{equation}
\vec{h}\approx\frac{K}{r^2}\left(\frac{\partial^2n_x}{\partial\phi^2}\hat{\mathbf{e}}_x+\frac{\partial^2n_y}{\partial\phi^2}\hat{\mathbf{e}}_y+\frac{\partial^2n_z}{\partial\phi^2}\hat{\mathbf{e}}_z\right),
\end{equation}
where $r$ is a radial distance from the defect line and $\phi$ is the azimuthal angle. For small enough $r$, the nematic is locally in equilibrium given by the condition $\vec{n}\times\vec{h}=0$, which is solved by the nbit form of the director field~\cite{BinyshJ_PhysRevLett124_2020}
\begin{equation}
\vec{n}(\phi)=-\sin\left(\frac{\phi}{2}+\gamma-\alpha\right)\vec{m}+\cos\left(\frac{\phi}{2}+\gamma-\alpha\right)\left(\cos\beta\vec{t}\times\vec{m}+\sin\beta\vec{t}\right).
\label{eq:spinor}
\end{equation}
Here, $\vec{t}$ is the defect line tangent which we take in the $z$ direction, $\vec{m}=(\cos(\alpha+\pi/2),\sin(\alpha+\pi/2),0)$, and $\alpha$, $\beta$, and $\gamma$ are arbitrary angles defining the nbit state; the domain of validity of this nbit solution is also discussed in Supplementary Figure S1.
As $\phi$ is increased from $0$ to $2\pi$, the director field traces half of a great circle on a unit sphere as shown in Fig.~2A  of the Main Text. In this work, we use quaternionic rotations of a reference director field defect profile to describe a whole range of solutions given by Eq.~\eqref{eq:spinor} and to construct a spinor description of nbits.

\section{Nbit dynamical equation}
The time evolution equation for nbit solutions can be derived from hydrodynamic models of nematic liquid crystals. Our derivation below is based on the Lagrangian formalism for the nematic free energy functional and the Rayleigh dissipation function $D$,  analogous to the derivations of dynamical equations for the $\pm 1/2$ defect motions~\cite{KlemanM} and flow-alignment~\cite{EmersicT_ScienceAdvances5_2019,TangX_PhysRevE101_2020} in nematics.

\subsubsection*{Dissipation function}
The Rayleigh dissipation function for slow director field deformations that generate only negligible velocity fields reads
\begin{equation}
D=\frac{\Gamma}{2}\left(\frac{\dd\vec{n}}{\dd t}\right)^2.
\label{eq:diss}
\end{equation}
We want to express the dissipation function in the Pauli algebra using the Pauli matrices
\begin{equation}
\sigma_x=\begin{pmatrix}
0 & 1\\
1 & 0
\end{pmatrix},\qquad
\sigma_y=\begin{pmatrix}
0 & \ii \\
-\ii & 0
\end{pmatrix},\qquad
\sigma_z=\begin{pmatrix}
1 & 0\\
0 & -1
\end{pmatrix}.
\end{equation}
Writing the director field in the Pauli algebra as $${\vec{n}}=n_x\sigma_x+n_y\sigma_y+n_z\sigma_z,$$  the dissipation function can be expressed as
\begin{equation}
D=\frac{\Gamma}{4}\Tr\left(\frac{\mathrm{d}{\vec{n}}}{\mathrm{d}t} \frac{\mathrm{d}{\vec{n}}}{\mathrm{d}t}\right).\label{eq:diss_Pauli}
\end{equation}
For the director field around a defect line, we take the nbit ansatz
\begin{equation}
{\vec{n}}=\eta{\vec{n}}_0\eta^\dagger,\label{eq:nbit_trans}
\end{equation}
where $\eta$ determines the quaternionic rotation of the reference profile 
\begin{equation}
{\vec{n}}_0=\cos\frac{\phi}{2}\sigma_x+\sin\frac{\phi}{2}\sigma_y.\label{eq:n0}
\end{equation}
Using Eq.~\eqref{eq:nbit_trans}, the dissipation function can be rewritten as
\begin{equation}
D=\frac{\Gamma}{4}\Tr\left\{\frac{\mathrm{d}}{\mathrm{d}t}\left[\eta\left(\sigma_x\cos\frac{\phi}{2}+\sigma_y\sin\frac{\phi}{2}\right)\eta^\dagger\right] \frac{\mathrm{d}}{\mathrm{d}t}\left[\eta\left(\sigma_x\cos\frac{\phi}{2}+\sigma_y\sin\frac{\phi}{2}\right)\eta^\dagger\right]\right\}.
\end{equation}
Note that $\eta$ can in principle depend on both the radial distance $r$ and the $z$ component, but $\eta$ has no angular dependence. To obtain the dynamics of $\eta$, we integrate the dissipation function over the azimuthal angle $\phi$:
\begin{equation}
\begin{split}
D_\phi&=\int_0^{2\pi}\mathrm{d}\phi\; D\\
&=\frac{\pi}{4}\Gamma\Tr\left(\frac{\dd \eta}{\dd t}\sigma_x\eta^\dagger\frac{\dd \eta}{\dd t}\sigma_x\eta^\dagger + 4\frac{\dd \eta}{\dd t}\frac{\dd \eta^\dagger}{\dd t}  + \eta\sigma_x\frac{\dd \eta^\dagger}{\dd t}\eta\sigma_x\frac{\dd \eta^\dagger}{\dd t} +\frac{\dd \eta}{\dd t}\sigma_y\eta^\dagger\frac{\dd \eta}{\dd t}\sigma_y\eta^\dagger + \eta\sigma_y\frac{\dd \eta^\dagger}{\dd t}\eta\sigma_y\frac{\dd \eta^\dagger}{\dd t} \right),
\end{split}
\label{eq:Dphi}
\end{equation}
where we have used the cyclic property of the trace and the fact that $\eta\eta^\dagger=\mathbb{1}$. We will also need the derivatives of the dissipation function with respect to $\dd \eta/\dd t$ and $\dd \eta^\dagger/\dd t$: 
\begin{align}
\left(\frac{\partial D_\phi}{\partial\left(\dd\eta/\dd t\right)}\right)^\intercal&=\frac{\pi}{2}\Gamma\left[ 2\frac{\dd\eta^\dagger}{\dd t}-\Tr\left(\eta^\dagger\frac{\dd\eta}{\dd t}\sigma_z\right)\sigma_z\eta^\dagger \right],\label{eq:Dt1}\\
\left(\frac{\partial D_\phi}{\partial\left(\dd\eta^\dagger/\dd t\right)}\right)^\intercal&=\frac{\pi}{2}\Gamma\left[ 2\frac{\dd\eta}{\dd t}-\Tr\left(\frac{\dd\eta^\dagger}{\dd t}\eta\sigma_z\right)\eta\sigma_z \right]=\left(\frac{\partial D_\phi}{\partial\left(\dd\eta/\dd t\right)}\right)^*,\label{eq:Dt2}
\end{align}
where we used the identity 
\begin{equation}
\left(\frac{\dd\eta}{\dd t}\right)^\dagger=\frac{\dd\eta^\dagger}{\dd t}
\end{equation}
and the fact that, for an arbitrary Pauli vector $\vec{p}=p_x\sigma_x+p_y\sigma_y+p_z\sigma_z$, 
\begin{equation}
\sigma_x\vec{p}\sigma_x+\sigma_y\vec{p}\sigma_y=-\Tr\left(\vec{p}\sigma_z\right)\sigma_z.\label{vector_identity}
\end{equation}
Furthermore, one finds from Eqs.~\eqref{eq:Dt1}~and~\eqref{eq:Dt2} that
\begin{equation}
\frac{\dd\eta}{\dd t}=\frac{1}{\pi\Gamma}\left\{ \left(\frac{\partial D_\phi}{\partial\left(\dd\eta^\dagger/\dd t\right)}\right)^\intercal - \frac{1}{4}\Tr\left[\left(\frac{\partial D_\phi}{\partial\left(\dd\eta^\dagger/\dd t\right)}\right)^\intercal\sigma_z\eta^\dagger\right]\eta\sigma_z  \right\}.
\label{eq:DetaDt}
\end{equation}
The derivation of Eq.~\eqref{eq:DetaDt} uses the identity $$\Tr\left[\left(\frac{\partial D_\phi}{\partial\left(\dd\eta^\dagger/\dd t\right)}\right)^\intercal\sigma_z\eta^\dagger\right]=-\Tr\left[\left(\frac{\partial D_\phi}{\partial\left(\dd\eta/\dd t\right)}\right)^\intercal\eta\sigma_z\right]$$  which follows from Eqs.~\eqref{eq:Dt1} and \eqref{eq:Dt2}.

\subsubsection*{Elastic free energy}
The elastic free energy in Eq.~\eqref{eq:free_energy} can be decomposed into derivatives w.r.t. $r$, $\phi$, and $z$. Derivatives w.r.t. $\phi$ describe the elastic penalty for a defect director field compared to a homogeneous director field and do not affect the nbit dynamics. Terms including the $z$ derivatives in the elastic free energy contribution $f_\text{el}^z \propto \left({\partial\vec{n}}/{\partial z}\right)^2$ can be written as
\begin{equation}
f_\text{el}^z=\frac{K}{4}\Tr\left( \frac{\partial\vec{n}}{\partial z}\frac{\partial\vec{n}}{\partial z}  \right).
\end{equation}
Following the same procedure as for Eq.~\eqref{eq:diss_Pauli}, $f_\text{el}^z$ can be integrated over $\phi$ and expressed as
\begin{equation}
f_{\text{el}\phi}^z=\frac{\pi}{4}K\Tr\left(\frac{\partial \eta}{\partial z}\sigma_x\eta^\dagger\frac{\partial \eta}{\partial z}\sigma_x\eta^\dagger + 4\frac{\partial \eta}{\partial z}\frac{\partial \eta^\dagger}{\partial z}  + \eta\sigma_x\frac{\partial \eta^\dagger}{\partial z}\eta\sigma_x\frac{\partial \eta^\dagger}{\partial z} +\frac{\partial \eta}{\partial z}\sigma_y\eta^\dagger\frac{\partial \eta}{\partial z}\sigma_y\eta^\dagger + \eta\sigma_y\frac{\partial \eta^\dagger}{\partial z}\eta\sigma_y\frac{\partial \eta^\dagger}{\partial z} \right)
\end{equation}
We will further need 
\begin{equation}
\left(\frac{\partial f_{\text{el}\phi}^z}{\partial\left(\partial\eta^\dagger/\partial z\right)}\right)^\intercal=\frac{\pi}{2}K\left( 2\frac{\partial\eta}{\partial z} +\eta\sigma_x\frac{\partial\eta^\dagger}{\partial z}\eta\sigma_x  +\eta\sigma_y\frac{\partial\eta^\dagger}{\partial z}\eta\sigma_y \right).
\end{equation}
Differentiating w.r.t. $z$, considering Eq.~\eqref{vector_identity} and the fact that 
$$\Tr\left(\frac{\partial\eta^\dagger}{\partial z}\frac{\partial\eta}{\partial z}\sigma_z\right)=0,$$
we obtain
\begin{equation}
\frac{\dd}{\dd z}\left(\frac{\partial f_{\text{el}\phi}^z}{\partial\left(\partial\eta^\dagger/\partial z\right)}\right)^\intercal=\frac{\pi}{2}K\left[ 2\frac{\partial^2\eta}{\partial z^2} -\Tr\left(\frac{\partial^2\eta^\dagger}{\partial z^2}\eta\sigma_z\right)\eta\sigma_z -\Tr\left(\frac{\partial\eta^\dagger}{\partial z}\eta\sigma_z\right)\frac{\partial\eta}{\partial z}\sigma_z  \right],
\end{equation}
which can be rewritten in the form
\begin{equation}
\frac{\dd}{\dd z}\left(\frac{\partial f_{\text{el}\phi}^z}{\partial\left(\partial\eta^\dagger/\partial z\right)}\right)^\intercal-\frac{1}{4}\Tr\left[\frac{\dd}{\dd z}\left(\frac{\partial f_{\text{el}\phi}^z}{\partial\left(\partial\eta^\dagger/\partial z\right)}\right)^\intercal\sigma_z\eta^\dagger\right]\eta\sigma_z=\pi K\frac{\partial^2\eta}{\partial z^2}-\frac{\pi K}{2}\Tr\left(\frac{\partial\eta^\dagger}{\partial z}\eta\sigma_z\right)\frac{\partial\eta}{\partial z}\sigma_z.
\label{eq:partial_z}
\end{equation}
The same derivation can be repeated for the radial nbit dependence, yielding
\begin{equation}
\frac{\dd}{\dd r}\left(\frac{\partial f_{\text{el}\phi}^z}{\partial\left(\partial\eta^\dagger/\partial r\right)}\right)^\intercal-\frac{1}{4}\Tr\left[\frac{\dd}{\dd r}\left(\frac{\partial f_{\text{el}\phi}^z}{\partial\left(\partial\eta^\dagger/\partial r\right)}\right)^\intercal\sigma_z\eta^\dagger\right]\eta\sigma_z=\pi K\frac{\partial^2\eta}{\partial r^2}-\frac{\pi K}{2}\Tr\left(\frac{\partial\eta^\dagger}{\partial r}\eta\sigma_z\right)\frac{\partial\eta}{\partial r}\sigma_z.
\label{eq:partial_r}
\end{equation}


\subsubsection*{Electric field}
The electric field contribution to the free energy in Eq.~\eqref{eq:free_energy} can be written as
\begin{equation}
f_E=-\frac{\epsilon_\text{a}}{8}\left[\Tr\left(\vec{n}\vec{E}\right)  \right]^2=-\frac{\epsilon_\text{a}}{8}\left[\Tr\left(\eta\vec{n}_0\eta^\dagger\vec{E}\right)  \right]^2,
\end{equation}
where the dielectric anisotropy $\epsilon_\text{a}$ is defined by $\epsilon_\text{a}=\varepsilon_\text{a}\varepsilon_0$~\cite{KlemanM}. 
In close proximity to the defect line, the director field remains well described by the nbit ansatz [Eq.~\eqref{eq:spinor}]; however, different nbit states can have different free energy due to the electric field contribution.
Using the reference profile from Eq.~\eqref{eq:n0} and integrating over azimuthal angle $\phi$, we obtain
\begin{align}
f_{E\phi}&=-\frac{\pi}{8}\epsilon_\text{a}\left\{ \left[\Tr\left(\eta\sigma_x\eta^\dagger\vec{E}\right)\right]^2 + \left[\Tr\left(\eta\sigma_y\eta^\dagger\vec{E}\right)\right]^2 \right\}\\
&=-\frac{\pi}{8}\epsilon_\text{a}\left\{|\vec{E}|^2-\left[\Tr\left(\eta\sigma_z\eta^\dagger\vec{E}\right)\right]^2\right\}.
\end{align}
The derivative of the electric free energy density w.r.t. $\eta^\dagger$ is given by
\begin{equation}
\left( \frac{\partial f_{E\phi}}{\partial\eta^\dagger} \right)^\intercal=\frac{\pi}{4}\epsilon_\text{a}\Tr\left(\eta\sigma_z\eta^\dagger\vec{E}\right)\vec{E}\eta\sigma_z.\label{eq:contribution_E}
\end{equation}

\subsubsection*{Final form of the dynamical equation}

The Euler-Lagrange equation for the nbit dynamics reads
\begin{equation}
\frac{\partial D_\phi}{\partial\left(\dd\eta^\dagger/\dd t\right)}+\frac{\delta f_\phi}{\delta\eta^\dagger}+\lambda\eta^\intercal=0,\label{eq:EL}
\end{equation}
where the Lagrange multiplier $\lambda$ preserves the SU(2) structure of $\eta$ and 
\begin{equation}
\frac{\delta f_\phi}{\delta\eta^\dagger}=\frac{\partial f_\phi}{\partial \eta^\dagger}-\frac{\dd}{\dd r}\frac{\partial f_\phi}{\partial\left(\partial\eta^\dagger/\partial r \right)}-\frac{\dd}{\dd z}\frac{\partial f_\phi}{\partial\left(\partial\eta^\dagger/\partial z \right)}.
\end{equation}
Inserting Eqs.~\eqref{eq:DetaDt}, \eqref{eq:partial_z}, \eqref{eq:partial_r}, and \eqref{eq:contribution_E} into Eq.~\eqref{eq:EL}, we obtain
\begin{equation}
\begin{split}
\frac{\dd\eta}{\dd t}&=\frac{K}{\Gamma}\left(\frac{\partial^2\eta}{\partial r^2} + \frac{\partial^2\eta}{\partial z^2}\right) -\frac{K}{2\Gamma}\Tr\left(\frac{\partial\eta^\dagger}{\partial z}\eta\sigma_z\right)\frac{\partial\eta}{\partial z}\sigma_z -\frac{K}{2\Gamma}\Tr\left(\frac{\partial\eta^\dagger}{\partial z}\eta\sigma_z\right)\frac{\partial\eta}{\partial z}\sigma_z\\
&\phantom{={}}-\frac{\epsilon_\text{a}}{4\Gamma}\Tr\left( \vec{E}\eta\sigma_z\eta^\dagger \right)\vec{E}\eta\sigma_z  -\frac{\lambda}{\pi\Gamma}\eta  ,
\label{eq:nbit1}
\end{split}
\end{equation}
The Lagrange multiplier
\begin{align}
\lambda&=\frac{1}{2}\Tr\left[\left(\frac{\delta f_\phi}{\delta\eta^\dagger}\right)^\intercal\eta^\dagger\right]
\label{eq:lambda2}
\end{align}
ensures that
$$\Tr\left[\left(\frac{\dd\eta}{\dd t}\right)\eta^\dagger\right]=0$$ 
and can be explicitly written as
\begin{equation}
\lambda=\frac{1}{2}\Tr\left\{\frac{K}{\Gamma}\left(\frac{\partial^2\eta}{\partial r^2} + \frac{\partial^2\eta}{\partial z^2}\right) -\frac{K}{2\Gamma}\Tr\left(\frac{\partial\eta^\dagger}{\partial z}\eta\sigma_z\right)\frac{\partial\eta}{\partial z}\sigma_z -\frac{K}{2\Gamma}\Tr\left(\frac{\partial\eta^\dagger}{\partial z}\eta\sigma_z\right)\frac{\partial\eta}{\partial z}\sigma_z
-\frac{\epsilon_\text{a}}{4\Gamma}\Tr\left( \vec{E}\eta\sigma_z\eta^\dagger \right)\vec{E}\eta\sigma_z \right\}.
\end{equation}

Equation~\eqref{eq:nbit1} governs the time dynamics of an nbit, with each term  having an obvious physical meaning:  The first term describes the elastic response due to nbit gradients in the radial and vertical direction. The expression $\eta\sigma_z\eta^\dagger$ appearing in the electric field term of Eq.~\eqref{eq:nbit1} is equal the director normal vector $\vec{\Omega}$ that describes the nbit orientation on a unit sphere. The term $\Tr\left( \vec{E}\eta\sigma_z\eta^\dagger \right)$ is proportional to $\vec{\Omega}\cdot\vec{E}$. After the Lagrange multiplier is applied, the term $\vec{E}\eta\sigma_z\eta^\dagger$ is proportional to $\vec{\Omega}\times\vec{E}$.
For the case $\epsilon_\text{a}<0$  considered in this paper, the electric field aims to align $\vec{\Omega}$ along $\vec{E}$ with the speed of alignment proportional to $|\epsilon_\text{a}\sin\left(2\phi\right)|$, where $\phi$ is the angle between $\vec{\Omega}$ and $\vec{E}$. 

The term 
\begin{equation}
\label{e:n_z^2}
-\frac{K}{2\Gamma}\Tr\left(\frac{\partial\eta^\dagger}{\partial z}\eta\sigma_z\right)\frac{\partial\eta}{\partial z}\sigma_z
\end{equation}
can be interpreted by writing the nbit derivative w.r.t. $z$ as ${\partial\eta}/{\partial z}=\mathrm{i}\vec{a}\eta$, where $\vec{a}$ is a non-normalised vector around which the local rotation of the nbit is performed.  We obtain
$$
-\frac{K}{2\Gamma}\Tr\left(\frac{\partial\eta^\dagger}{\partial z}\eta\sigma_z\right)\frac{\partial\eta}{\partial z}\sigma_z=-\frac{K}{2\Gamma}\text{Tr}\left(\vec{a}\eta\sigma_z\eta^\dagger\right)\vec{a}\eta\sigma_z
$$
which has the same structure as the electric field term in the second line of \eqref{eq:nbit1}. However, for $\epsilon_\text{a}<0$ the sign of the term~\eqref{e:n_z^2} is opposite to the sign of the electric field term and thus aims to align $\vec{\Omega}$ perpendicular to the rotation vector $\vec{a}$.  The same interpretation can be made for the associated term with the radial derivative.

\par
The discussion in the Main Text focusses on the regime of relatively strong electric fields with 
$$\frac{\epsilon_\text{a}|\vec{E}|^2L^2}{K}\gg 1,$$  where $L$ is the system size; in this case,  
the term~\eqref{e:n_z^2} becomes dominated by the electric field term in Eq.~\eqref{eq:nbit1}. Indeed, 
for the parameters in the Main Text, our numerical simulation of  Eq.~\eqref{eq:nbit1} confirm that the term~\eqref{e:n_z^2} can be neglected.
Therefore,  radially constant single-nbit solutions are governed by
\begin{eqnarray}
\frac{\dd\eta}{\dd t}=
\frac{K}{\Gamma} \frac{\partial^2\eta}{\partial z^2} -
\frac{\epsilon_\text{a}}{4\Gamma}\Tr\left( \vec{E}\eta\sigma_z\eta^\dagger \right)\vec{E}\eta\sigma_z  -\tilde{\lambda}\eta, 
\qquad\qquad
\tilde{\lambda}=\frac{\lambda}{\pi\Gamma}
\label{e:reduced}
\end{eqnarray}
which corresponds to Eq.~(5) of the Main Text (with tildes dropped).

\par
The dynamics of other observables, such as $\vec{\Omega}=\eta\sigma_z\eta^\dagger$, follows from Eq.~\eqref{eq:nbit1}. Finally, we also note that the nbit dynamics can be also  generalized to include effects of weak anisotropy of elastic deformation modes, weak chirality, magnetic fields, and defect line curvature.

\section{Energetic costs of nbit manipulation}
In this section, we calculate the dissipated energy as a $|0)$ nbit is transformed into a $|1)$ nbit. Initial configuration of the $|0)$ nbit is a $+1/2$ nematic defect aligned along the $x$-axis
\begin{equation}
\vec{n}=(\cos\frac{\phi}{2},\sin\frac{\phi}{2},0),
\end{equation}
where $\phi$ is the azimuthal angle. We perform the transformation by rotating the director field locally around the $y$-axis by an angle $\alpha$, obtaining
\begin{equation}
\frac{\mathrm{d}\vec{n}}{\mathrm{d}t}=-\frac{\mathrm{d}\alpha}{\mathrm{d}t}\cos\frac{\phi}{2}(\sin\alpha,0,\cos\alpha),
\end{equation}
The dissipated energy $E$ per defect line segment $L$ equals
\begin{align}
\frac{E}{L}&=\int_{R_\text{min}}^{R_\text{max}}r\mathrm{d}r\int_0^{2\pi}\mathrm{d}\phi\int_0^\tau\mathrm{d}t\,D\\
&=\int_{R_\text{min}}^{R_\text{max}}r\mathrm{d}r\int_0^{2\pi}\mathrm{d}\phi\int_0^\tau\mathrm{d}t\,\frac{\Gamma}{2}\left(\frac{\mathrm{d}\vec{n}}{\mathrm{d}t}\right)^2\\
&=\int_{R_\text{min}}^{R_\text{max}}r\mathrm{d}r\int_0^{2\pi}\mathrm{d}\phi\int_0^\tau\mathrm{d}t\,\frac{\Gamma}{2}\left(\frac{\mathrm{d}\alpha}{\mathrm{d}t}\right)^2\cos^2\frac{\phi}{2},
\end{align}
where $R_\text{min}$ and $R_\text{max}$ are the radial bounds of the defect region, $\tau$ is the time of the transformation, and $D$ is the dissipation function from Eq.~\eqref{eq:diss}. We take a constant rate of director rotation with $\frac{\mathrm{d}\alpha}{\mathrm{d}t}=\frac{\pi}{\tau}$. Also, we take $R_\text{min}\rightarrow 0$ and use the notation $R_\text{max}=R$. 
The final result for the dissipated energy equals
\begin{equation}
E=\frac{\pi^3\Gamma}{4\tau}LR^2.
\end{equation}

\section{Nematic Deutsch algorithm}

Deutsch algorithms~\cite{1997DeutschJozsa,PhysRevA.58.R1633,CleveR_Proceedings454_1998} played a conceptually important role in the development of quantum computation by demonstrating that certain problems  can be solved exponentially faster than with classical digital computation. Broadly, Deutsch algorithms aim to  determine global properties of Boolean functions $f:\{0,1\}^n\to \{0,1\}$ by using the smallest number of queries. The full entanglement-assisted power of quantum Deutsch algorithms  comes into play for $n>2$~\cite{PhysRevA.58.R1633}, and exponential speed-ups should not necessarily be expected in nematic systems. Nonetheless, considering  the elementary case $n=1$ is useful to illustrate the differences between nbit-computations and classical digital computations.  The specific goal is to use a single query to determine whether an unknown Boolean function  $f:\{0,1\}\to \{0,1\}$ is balanced, $f(0)\neq f(1)$, or constant, $f(0)=f(1)$. Building on the nbit representations and nematic logic gates, we consider the logic circuit in Fig.~S7, which presents the nematic analog of the quantum Deutsch algorithm in Fig.~3(b) of Ref.~\cite{CleveR_Proceedings454_1998}. The computation starts from the initial two-nbit product state $|\eta_1)\otimes |\eta_2)=|0)\otimes[|0)-|1)]/\sqrt{2}$.  In the first step, the phase-shifted Hadamard gate (Figs.~4D and S7) is applied to the first nbit $|\eta_1)=|0)$ to produce the superposition $|\eta_1')=[|0)+|1)]/\sqrt{2}$. As in the quantum case~\cite{CleveR_Proceedings454_1998}, it is assumed that the unknown Boolean black box function $f$ acts as a \quotes{f-controlled-NOT} defined by $|x)\otimes |y)\xrightarrow{f}|x)\otimes |y\oplus f(x))$ for $x,y\in \{0,1\}$, where \quotes{$\oplus$} represents addition modulo~2. 
Applying the defining~\cite{CleveR_Proceedings454_1998,PhysRevA.58.R1633} relations for \quotes{f-controlled-NOT} to the product states $|0) \otimes [|0)-|1)]$ and $|1) \otimes [|0)-|1)]$ gives
\begin{eqnarray}
 |0) \otimes [|0\oplus f(0))-|1\oplus f(0))] =(-1)^{f(0)}  |0)\otimes [|0)-|1)] ,
 \notag\\
 |1) \otimes [|0\oplus f(1))-|1\oplus f(1))] =(-1)^{f(1)}  |1)\otimes [|0)-|1)] .
 \notag
\end{eqnarray}
Omitting the normalization factor~$1/2$, \quotes{f-controlled-NOT} thus transforms the pre-black box state $[|0)+|1)] \otimes [|0)-|1)]$ into the post-black box state
\begin{equation}
 [(-1)^{f(0)}|0)+(-1)^{f(1)}|1)] \otimes [|0)-|1)].
 \label{e:fcN}
\end{equation}
The auxiliary second nbit remains unchanged throughout. Finally, applying a second phase-shifted Hadamard gate to the first nbit, the algorithm returns for the first nbit
\begin{equation}
(-1)^{f(0)} | f(0)\oplus f(1) )
\end{equation}
corresponding to a $+1/2$ defect state $\pm|0)$ when $f$ is constant, or a $-1/2$ defect state $\pm|1)$  when $f$ is balanced (Fig.~S7). Thus, by exploiting single-nbit superposition,  the nematic Deutsch algorithm can determine a global property of the black box function from a single run. Note that, for  the specified initial state, all operations involved only two-nbit product states and hence can be implemented using the concepts developed above. 
To summarize, although many quantum algorithms are unlikely to  permit nematic counterparts of comparable complexity, the example in Fig.~S7  suggests that suitably posed problems can be solved leveraging nbit superpositions.

\newpage

\section{Supplementary figures}
{
\centering
\includegraphics[width=15 cm]{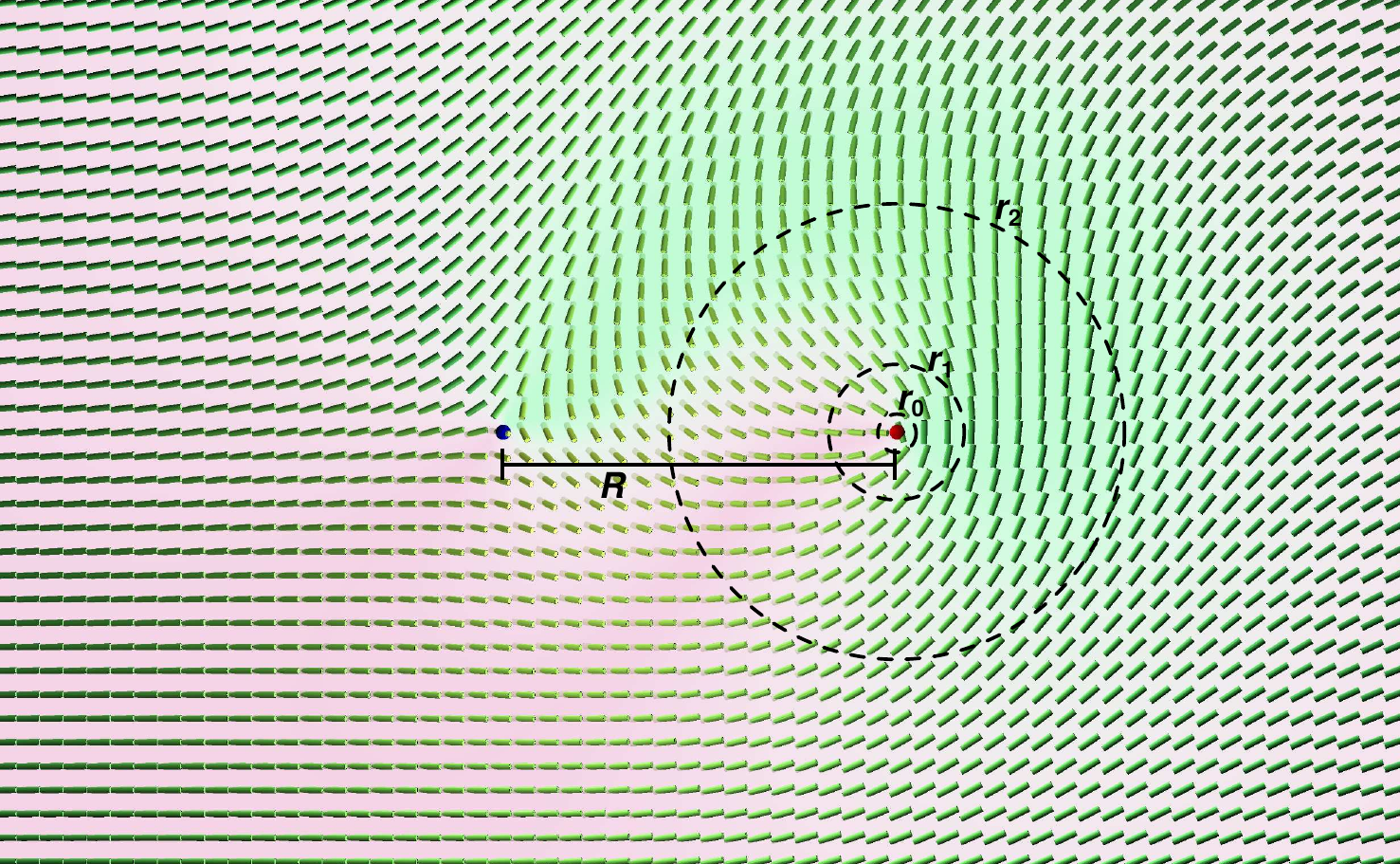}
\\
\vspace*{10pt}
{\small Supplementary Figure S1: Non-equilibrium director field around two half-integer nematic defect lines. The nbit solution~\eqref{eq:spinor} is valid in close proximity to the defect ($r_0\ll R$). The director profile deviates from the nbit profile as one moves  radially away from the defect line; however,  at small distances from the defect line ($r\sim r_1$), the director field is still close to an nbit form.
By contrast, at larger distances ($r\sim r_2$), the director field profile has to be computed from the full complete time-dependent director field dynamics [Eq.~\eqref{eq:field}].}
}

\newpage

{
\centering
\includegraphics[width=\textwidth]{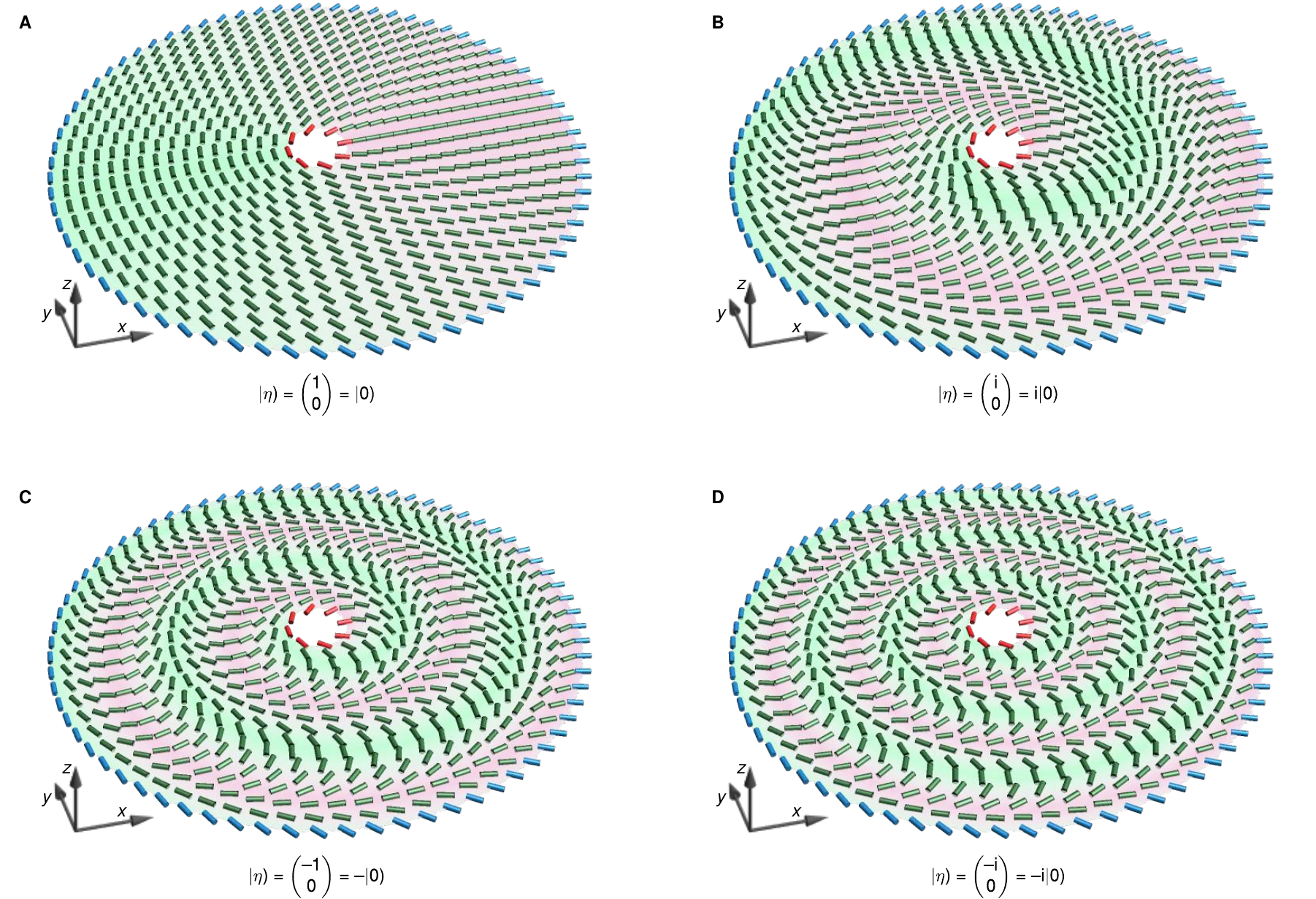}
\\
\vspace*{10pt}
{\small Supplementary Figure S2: Four locally equivalent configurations of the $|0)$ nbit profile. The configurations are obtained by rotating the inner director field (red rods) around the $z$-axis while the reference director field (outer blue rods) remains fixed. For all four solutions, the inner red director field is the same, but the intermediate director field between red and blue regions is distinctively different and cannot be smoothly transformed from one configuration to another provided that inner and outer director field are kept fixed. The degree of rotation of the inner red director field equals (A) $0$, (B) $\pi$, (C) $2\pi$, and (D) $3\pi$. A $4\pi$-rotation can be smoothly transformed into $|0)$ nbit profile, as demonstrated in Supplementary Movie 2.}
}

\newpage

{
\centering
\includegraphics[width=\textwidth]{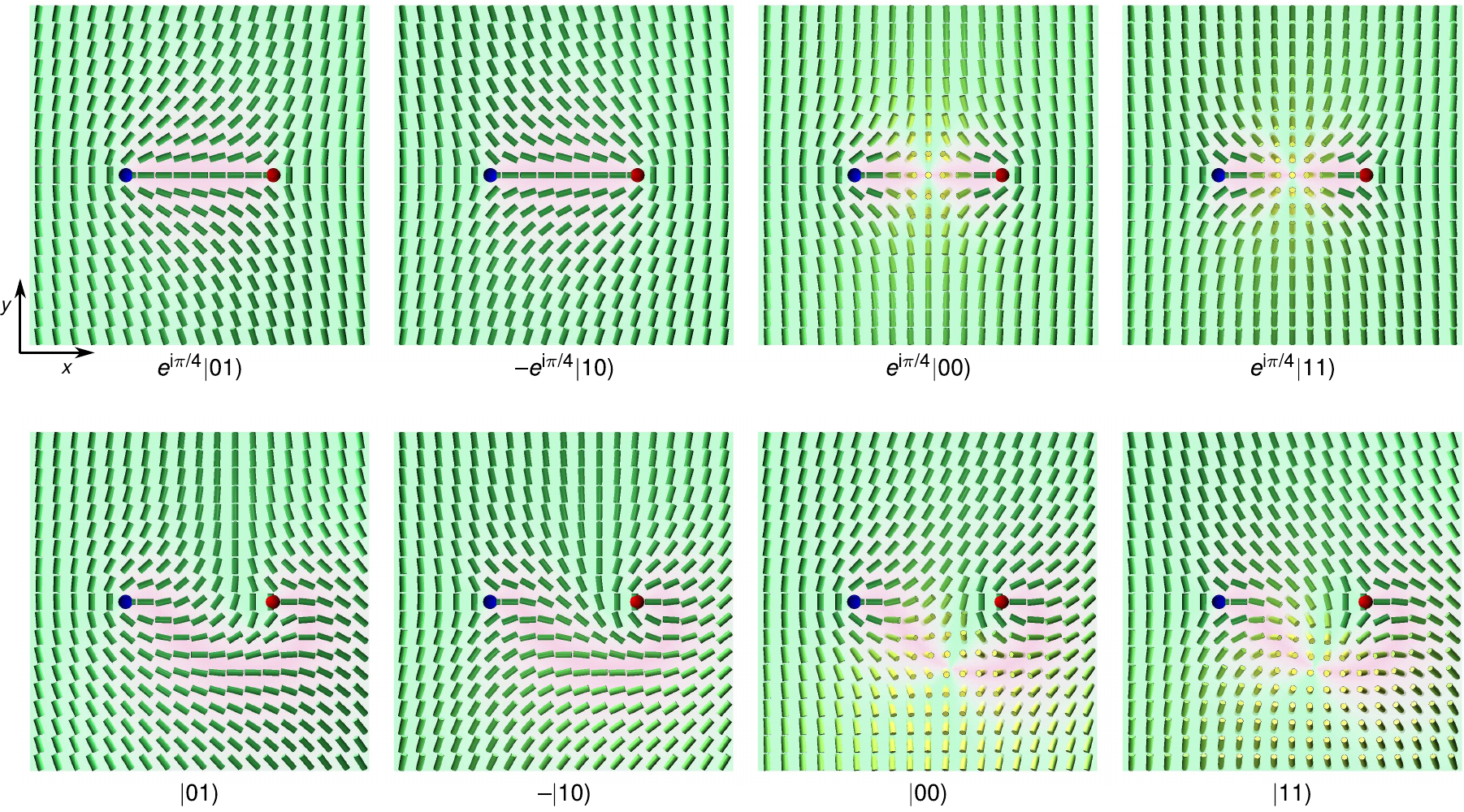}
\\
\vspace*{10pt}
{\small Supplementary Figure S3: Four product basis states with a global phase (top row) and without a global phase (bottom row), embedded in a director far field along $y$ direction. Only the states $e^{\mathrm{i}\pi/4}|01)$ and $-e^{\mathrm{i}\pi/4}|10)$  correspond to minima of the free energy. See also  Fig.~S5.}
}

\noindent\begin{minipage}{\textwidth}
{
\centering
\includegraphics[]{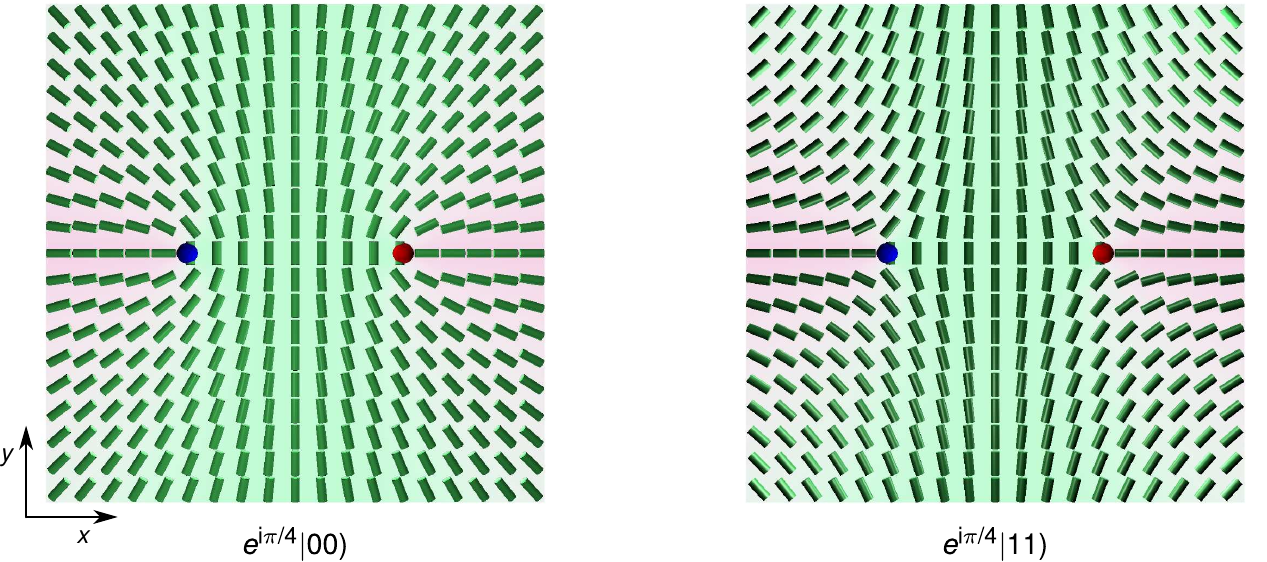}
\\
\vspace*{10pt}
{\small Supplementary Figure S4:  Alternative realization of the two-nbit states $e^{\mathrm{i}\pi/4}|00)$ and $e^{\mathrm{i}\pi/4}|11)$ without an umbilic soliton. Compared to Fig.~S3, the umbilic has been moved towards infinity along the $y$ axis.}
}
\end{minipage}

 \noindent\begin{minipage}{\textwidth}
{
\centering
\includegraphics[]{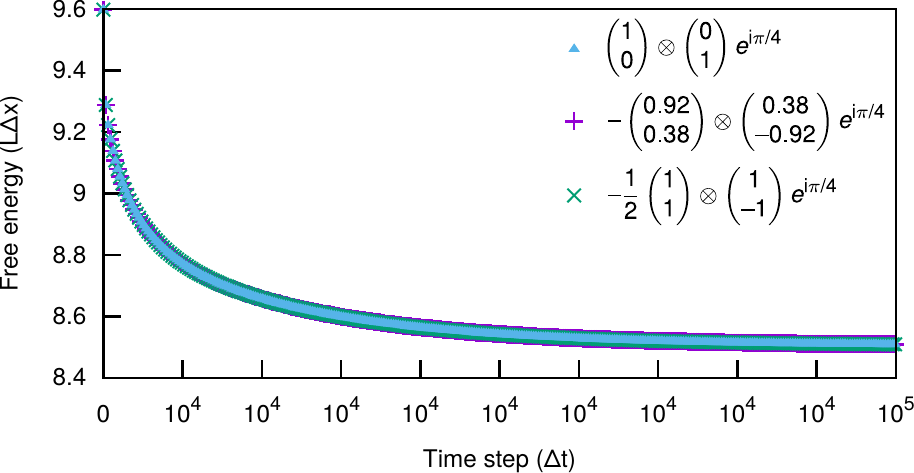}
\\
\vspace*{10pt}
{\small Supplementary Figure S5: Energetic equivalence of two-nbit states in the $\Psi^-$ ensemble. We show the free energy relaxation for 3 states from the $\Psi^-$ ensemble manifold that are also shown in Fig.~5B in the Main Text. The initial condition in the left and right half of the simulation plane corresponds to the director ansatz [Eq.~\eqref{eq:spinor}] for the left and right defect, respectively. Simulation was performed for periodic boundary  conditions,  using gradient descent as explained in Methods. Not only have the states equal final free energy, but they also follow the same relaxation curve. The same dependency is obtained for states from the $\Psi^-$ ensemble manifold that have a far director field in an arbitrary direction.}
}
\end{minipage}

 \noindent\begin{minipage}{\textwidth}
{
\centering
\includegraphics[]{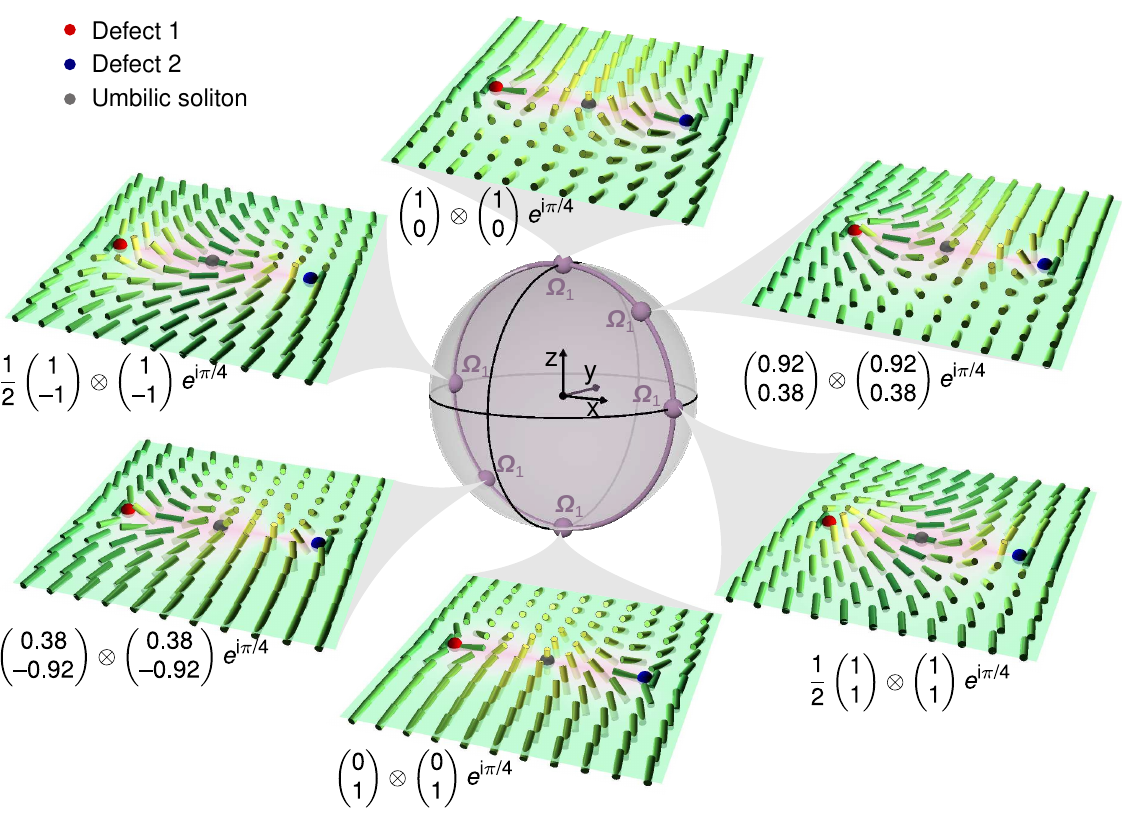}
\\
\vspace*{10pt}
{\small Supplementary Figure S6: Representative examples  from the ensemble manifold $\Phi^+$ of parallel two-nbit states  with $\vec{\Omega}_1=\vec{\Omega}_2$. The examples are energetically equivalent, with the director far-field parallel to the $y$ direction. The states in the manifold are made stable by enforcing an umbilic solition at the center of the system.}
}
\end{minipage}

 \noindent\begin{minipage}{\textwidth}
{
\centering
\includegraphics[]{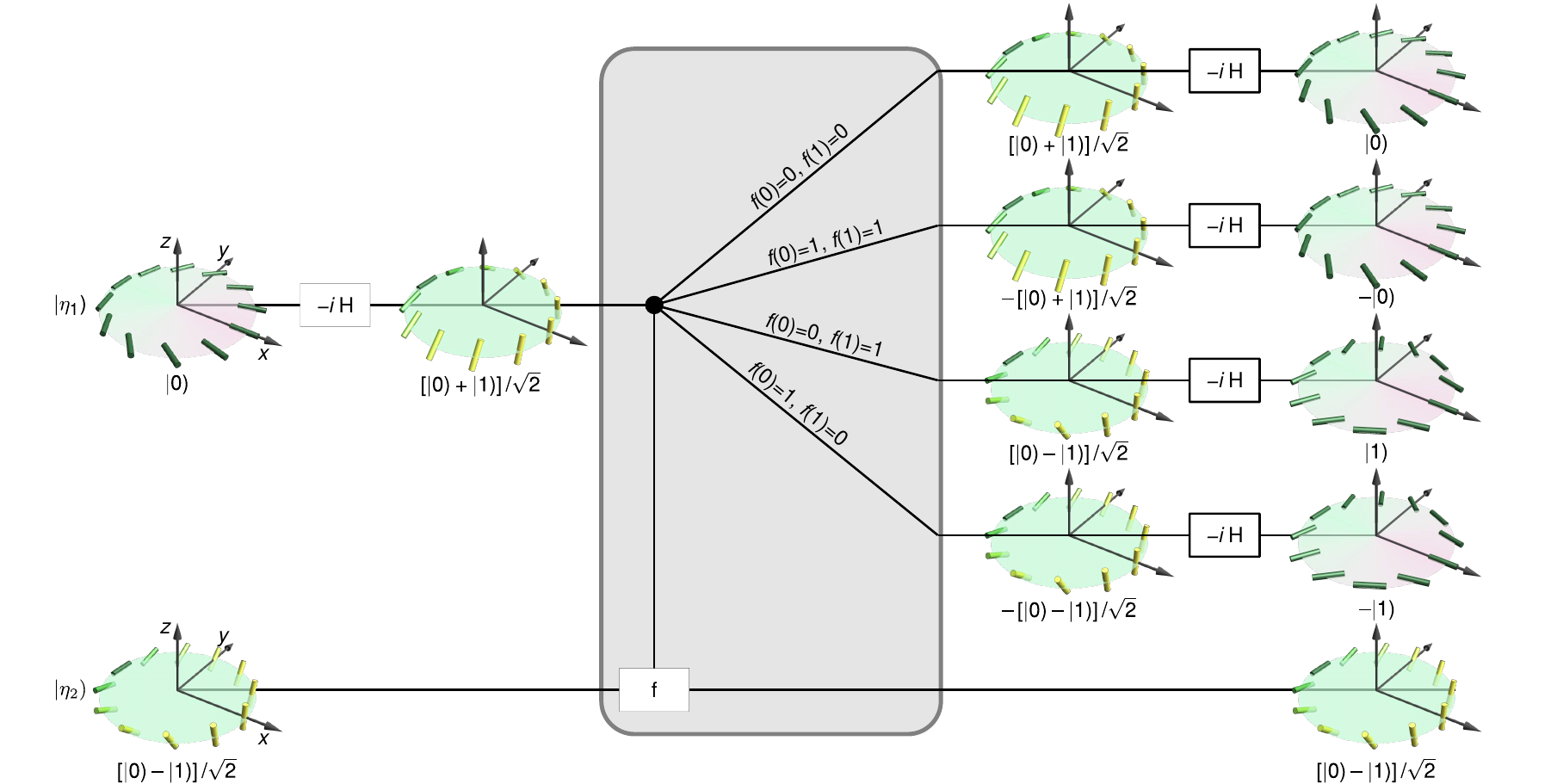}
\\
\vspace*{10pt}
{\small Supplementary Figure S7: Logical circuit for a single-nbit Deutsch algorithm.
The goal is to determine, from a single query, whether an unknown Boolean \lq black box\rq{} function $f:\{0,1\}\to\{0,1\}$ is constant [$f(0)=f(1)$]  or balanced [$f(0)\neq f(1)$]. The input of the algorithm is the two-nbit product state $|\eta_1)\otimes |\eta_2)=|0)\otimes [|0)-|1)]/\sqrt{2}$.  The auxiliary~\cite{PhysRevA.58.R1633} nbit $|\eta_2)$ remains unchanged throughout the computation. First, a phase-shifted Hadamard gate (Fig.~4D) 
$-i\mathrm{H}$ 
is applied to the first nbit, where the phase factor $-i$ is obtained by rotating each director around $\vec{\Omega}$ by an angle  $-\pi$ (Fig.~2A). Next, a black box two-nbit operation of \lq f-controlled-NOT\rq{} is performed on both nbits~\cite{CleveR_Proceedings454_1998}, changing the first nbit but not the second [Eq.~\eqref{e:fcN}].  There exist four possible Boolean functions $f:\{0,1\}\to\{0,1\}$, each giving a different outcome for the first nbit. Upon application of a second Hadamard gate to the first nbit, the circuit will return the first nbit  in a +1/2-defect state $\pm|0)$ if $f$ is constant, or in a $-1/2$-defect state $\pm|1)$ if $f$ is balanced.
}
}
\end{minipage}

\section{Supplementary movies}

{
\centering
\includegraphics[width=0.7\textwidth]{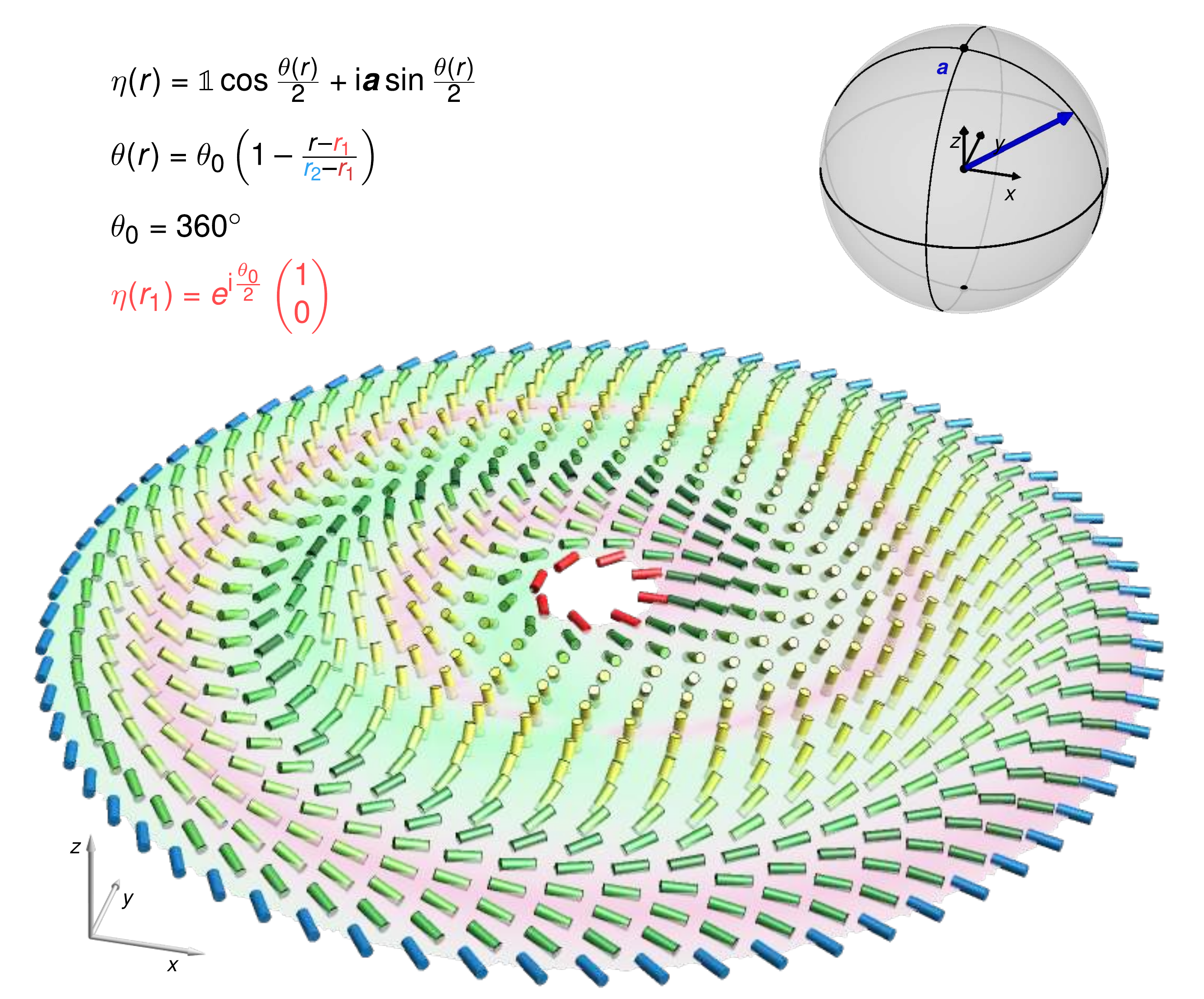}
\\
\vspace*{10pt}
{\small Supplementary Movie 1: Creation and manipulation of the $-|0)$ nbit. During the first stage of the protocol, the $-|0)$ nbit is created by rotating the red director field (at the smallest radius shown) by $360^\circ$ around $z$-axis, relative to the blue director field (at maximum radius) that is kept fixed in the reference configuration. The angle of rotation changes linearly with decreasing radius. Once the $360^\circ$ rotation has been completed, both red and blue director fields are kept fixed. During the subsequent second stage,  we change the orientation of the axis, around which the director field is rotated from the reference profile (blue) inwards. The video demonstrates how rotating the director field of a $|0)$ profile by $360^\circ$ results in a $-|0)$ profile,  regardless of the chosen axis of rotation. The director deformation between $|0)$ and $-|0)$ profile can therefore be smoothly transformed between clockwise spiral to anticlockwise spiral. Movie is available upon request.}
}

\newpage

{
\centering
\includegraphics[width=0.7\textwidth]{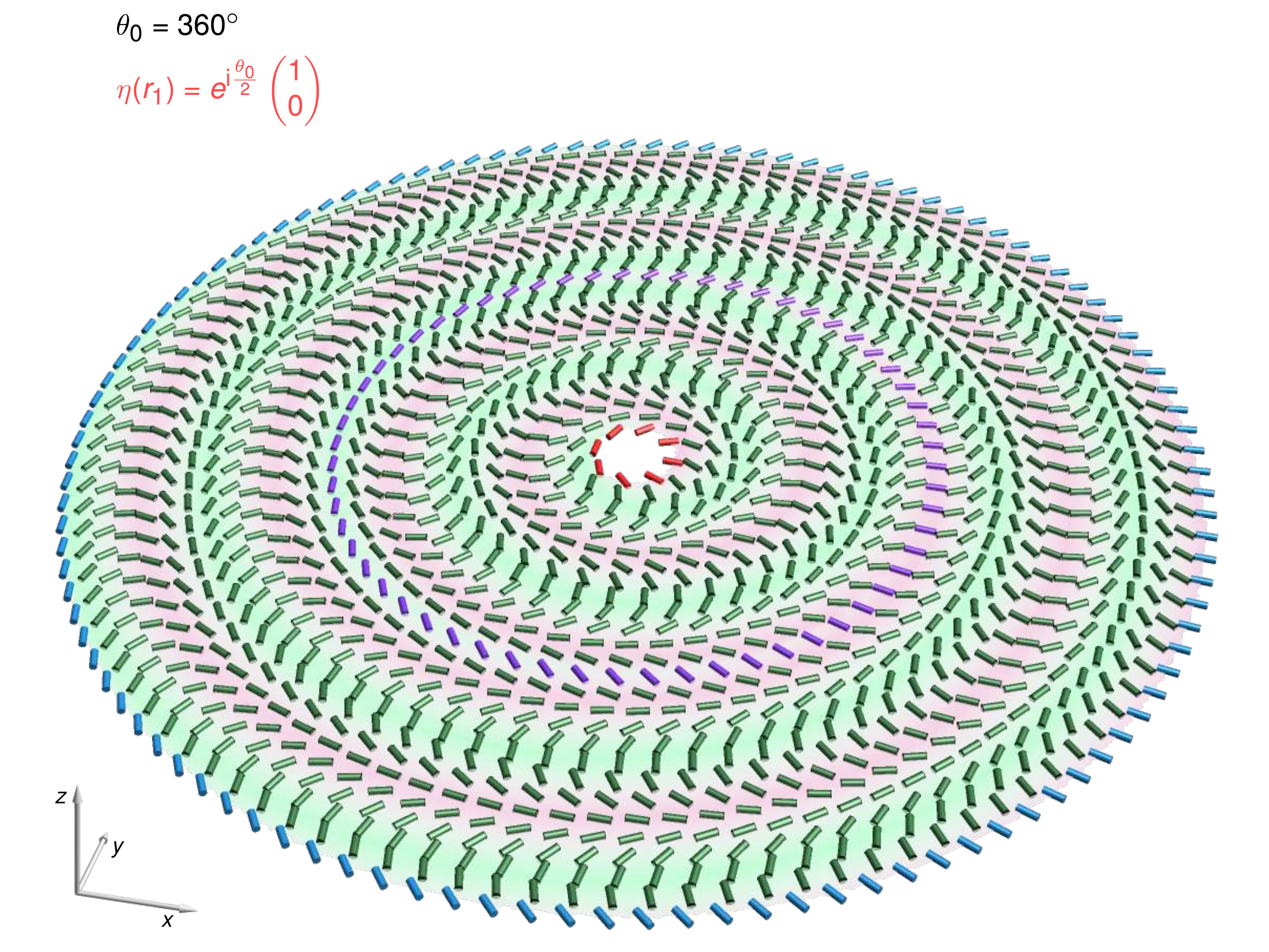}
\\
\vspace*{10pt}
{\small Supplementary Movie 2: Nbit rotation by $720^\circ$  is realized in three steps. Step~1: In the first part of the video, the inner red director field is rotated by $720^\circ$ relative to the outer blue reference profile. The intermediate  purple  director profile gets rotated by $360^\circ$ and is in the $-|0)$ configuration. According to Eq.~(2) in the Main Text, the red director profile should now have returned to the $|0)$ configuration, since $720^\circ$ rotation does not change a spinor. 
We show this explicitly by performing, in steps 2 and 3, a smooth reconfiguration of the director field that produces a homogeneous $+1/2$ defect director profile, while keeping the inner red and the outer blue director profiles fixed. 
Step~2:  To achieve such a reconfiguration, we reverse the direction of the director rotation between the purple and the blue ring  in the second part of the video; this process is explained in detail in Supplementary Movie 1. At the end of part 2, the director field changes from clockwise to anticlockwise rotation as the radius is increased, leading to a transition from an inner right-handed to the outer left-handed spiral pattern in the phase field. Step~3: In the third part, the mismatch between clockwise and anticlockwise rotation relaxes into a homogeneous $+1/2$ director field profile. Together with Supplementary Movie~1, this example shows explicitly how a rotation of the director field by $360^\circ$ changes the sign of the spinor, whereas a  rotation by $720^\circ$ has no effect on the spinor form. Movie is available upon request.}
}

\newpage

{
\centering
\includegraphics[width=\textwidth]{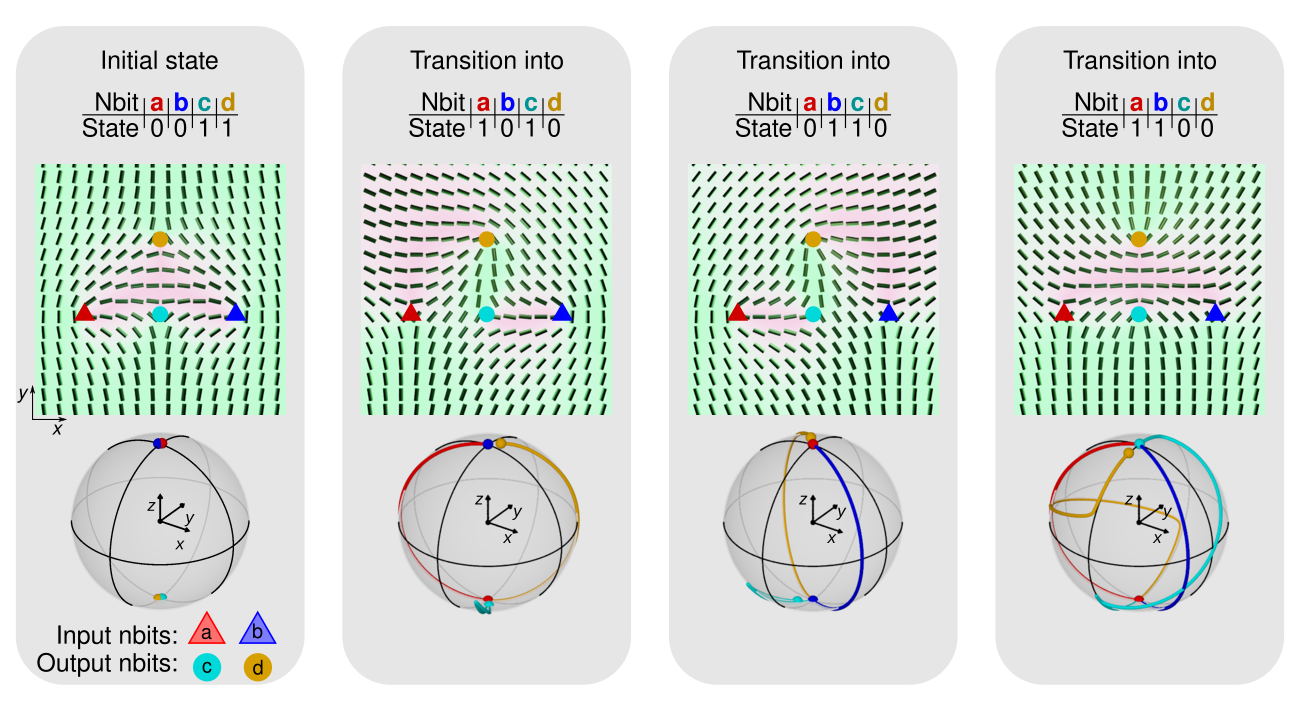}
\\
\vspace*{10pt}
\small Supplementary Movie 3: Universal classical logic gates. The movie shows director field reconfiguration and nbit states on the Bloch sphere corresponding to transformations in Fig.~6. First panel shows the initial state of two input nbits and two output nbits. To apply the NAND and NOR logical operation, either one or both of the input nbits `a' and `b' are flipped in the second, third and fourth panel, and the response of the output nbits `c' and `d' is observed. The movie duration is $2\cdot 10^5\,\Delta x^2/(\Gamma L)$, where $\Delta x$ is the mesh resolution and the distance between `a' and `c' nbits equals $50\,\Delta x$. Movie is available upon request.
}

\end{document}